\documentclass[apj]{emulateapj}
\usepackage{mathtools}
\usepackage{natbib}
\usepackage[breaklinks,colorlinks,urlcolor=blue,citecolor=blue,linkcolor=blue]{hyperref}
\usepackage{color}
\usepackage{enumitem}
\usepackage{graphicx,xspace,color,longtable}
\usepackage{amssymb,amsmath,shadow,bezier,curves,rotating}
\usepackage[toc,titletoc]{appendix}
\usepackage[flushleft]{threeparttable}

\shorttitle{Cosmological Constraint on \om ~and \sig ~from $\mathtt{GalWCat19}$ Cluster Catalog}
\shortauthors{Abdullah et al. 2020}

\newcommand {\h} {$h^{-1}\,$Mpc}
\newcommand {\ks} {km s$^{-1}$}
\newcommand {\hm} {$h^{-1} \  M_{\odot}$}

\newcommand {\sig} {$\sigma_8$}
\newcommand {\om} {$\Omega_m$}

\begin{document}

\title{Cosmological Constraints on \om ~and \sig\ from Cluster Abundances using the  $\mathtt{GalWCat19}$ Optical-Spectroscopic SDSS Catalog}

\author{Mohamed H. Abdullah}
\affil{Department of Physics and Astronomy, University of California Riverside, 900 University Avenue, Riverside, CA 92521, USA\\
       Department of Astronomy, National Research Institute of Astronomy and Geophysics, Helwan, 11421, Egypt}
\email{melha004@ucr.edu}

\author{Anatoly Klypin}
\affil{Astronomy Department, New Mexico State University, Las Cruces, NM 88001, USA
\\
Department of Astronomy, University  of Virginia, Charlottesville, VA 22904, USA}

\author{Gillian Wilson}
\affil{Department of Physics and Astronomy, University of California Riverside, 900 University Avenue, Riverside, CA 92521, USA}

\begin{abstract} 
We derive cosmological constraints on the matter density, \om, and the  amplitude of fluctuations, \sig, using $\mathtt{GalWCat19}$, a  catalog  of 1800 galaxy clusters we identified  in the Sloan Digital Sky Survey-DR13 spectroscopic data set using our GalWeight technique to determine cluster membership \citep{Abdullah18,Abdullah19}. By analyzing a subsample of 756 clusters in a redshift range of $0.045\leq z \leq 0.125$ and virial masses of $M\geq 0.8\times10^{14}$ \hm ~with mean redshift of $z = 0.085$, we obtain \om ~$=0.310^{+0.023}_{-0.027} \pm 0.041$ (systematic) and \sig ~$=0.810^{+0.031}_{-0.036}\pm 0.035$ (systematic), with a cluster normalization relation of  $\sigma_8= 0.43 \Omega_m^{-0.55}$. There are several unique aspects to our approach: we use the largest spectroscopic data set currently available, and we assign membership using the GalWeight technique which we have shown to be very effective at simultaneously maximizing the number of {\it{bona fide}} cluster members while minimizing the number of contaminating interlopers. Moreover, rather than employing scaling relations, we calculate cluster masses individually using the virial mass estimator. Since $\mathtt{GalWCat19}$ is a low-redshift cluster catalog we do not need to make any assumptions about evolution either in cosmological parameters or in the properties of the clusters themselves.  Our constraints on \om ~and \sig ~are consistent and very competitive with those obtained from non-cluster abundance cosmological probes such as Cosmic Microwave Background (CMB), Baryonic Acoustic Oscillation (BAO), and supernovae (SNe). The joint analysis of our cluster data with Planck18+BAO+Pantheon gives \om ~$=0.315^{+0.013}_{-0.011}$ and \sig ~$=0.810^{+0.011}_{-0.010}$.
\end{abstract}


\section{Introduction}
In the current picture of structure formation, galaxy clusters arise from rare high peaks of the initial density fluctuation field. These peaks grow in a hierarchical fashion through the dissipationless mechanism of gravitational instability with more massive halos growing via continued accretion and merging of low-mass halos \citep{White91,Kauffmann99,Kauffmann03}. Galaxy clusters are the most massive virialized systems in the universe and are uniquely powerful cosmological probes. The cluster mass function (CMF), or the abundance of galaxy clusters, is particularly sensitive to the matter density of the universe \om ~and \sig, the root-mean-square (rms) mass fluctuation on the scale of 8 \h ~at z = 0 (e.g., \citealp{  Wang98,Battye03,Dahle06,Wen10}). 

Cosmological analyses have been performed using samples of galaxy cluster constructed from galaxy surveys (e.g., \citealp{Rozo10,Kirby19,Abbott20}), X-ray emission (e.g., \citealp{Vikhlinin09,Mantz15}), and thermal Sunyaev-Zel'dovich (SZ) signal (e.g., \citealp{Bocquet19,Zubeldia19}). These cluster abundance studies showed that \om ~varies from $\sim$ 0.2 to 0.4 and \sig ~varies from $\sim$ 0.6 to 1.0. The discrepancies or tensions among these various studies is basically dependent on the accuracy of cluster mass estimation. Cluster mass can be calculated from cluster dynamics using, for example, the virial mass estimator (e.g., \citealp{Binney87}), the weak gravitational lensing \citep{Wilson96,Holhjem09}, and the application of Jeans equation for the gas density calculated from the x-ray analysis of galaxy cluster \citep{Sarazin88}. It can be also estimated from other observables, the so-called mass proxies, which scale tightly with cluster mass, such as X-ray luminosity (e.g., \citealp{Pratt09}), optical luminosity or richness (e.g. \citealp{Yee03,Simet17}), and the velocity dispersion of member galaxies (e.g., \citealp{Biviano06,Bocquet15}). Generally, most of these methods introduce large systematic uncertainties which limits the accuracy of estimating cluster masses (e.g., \citealp{Wojtak07,Mantz16a}). 

Cosmological analyses of galaxy cluster abundance introduce a degeneracy between \om ~and \sig. Large ongoing and upcoming wide and deep-field imaging and spectroscopic surveys at different redshifts, such as DES \citep{Abbott18a}, eROSITA \citep{Merloni12}, LSST \citep{LSST09}, and WFIRST \citep{Akeson19}, will simultaneously increase the precision of measuring the cosmological parameters and break the degeneracy between them. This is because \om ~evolves slowly while \sig ~evolves strongly with redshift. Also, these galaxy surveys at different redshifts are significant to study the evolution of the CMF which is critical to measuring structure growth, and therefore can be used to constrain properties of dark energy (e.g, \citealp{Haiman01,Mantz08}). Introducing advanced methods is essential to analyze these surveys. One of these methods is the GalWeight technique (\citealp{Abdullah18}, hereafter Abdullah+18) which can by applied to the available and upcoming spectroscopic database of eBOSS \citep{Raichoor17}, DESI \citep{Levi19}, and Euclid \citep{Euclid19III} to construct cluster catalogs. These catalogs provide an unlimited data source for a wide range of astrophysical and cosmological applications.

In addition, there are independent cosmological probes to constraining the cosmological parameters that can be applied alongside or in combination with galaxy cluster abundance. The anisotropies in the cosmic microwave background (CMB) are an independent probe of cosmological parameters (e.g., \citealp{Hinshaw13,Planck15}). The likelihoods of the \om-\sig ~confidence levels introduced by the CMF and CMB  are almost orthogonal to each other, which means combining these measurements will eliminate the degeneracy between \om  ~and \sig ~and shrink the uncertainties. Other independent cosmological probes that are used to constrain \om ~and \sig ~include cosmic shear, galaxy-galaxy lensing, and angular clustering (e.g, \citealp{Abbott18b,Uitert18}). The likelihoods of the \om-\sig ~confidence levels introduced by these probes are almost parallel to those introduced by the CMF. Moreover, the two cosmological probes of baryon acoustic oscillations (BAO, e.g., \citealp{Eisenstein05}) and supernovae (SNe, e.g., \citealp{Perlmutter99}) can be used to constrain \om ~only (independent of \sig). 

In this paper, we aim to derive the CMF and the cosmological parameters \om ~and \sig ~using a subsample of 756 clusters ($\mathtt{SelFMC}$) obtained from the $\mathtt{GalWCat19}$ cluster catalog as we discuss below in detail. The $\mathtt{GalWCat19}$ (\citealp{Abdullah19}, hereafter Abdullah+20) catalog was derived from the Sloan Digital Sky Survey-Data Release 13 spectroscopic data set (hereafter SDSS-DR13\footnote{\url{https://www.sdss.org/dr13/}}, \citealp{Albareti17}). The clusters were first identified by looking for the Finger-of-God effect (see, \citealp{Jackson72,Kaiser87,Abdullah13}). The cluster membership was constructed by applying our own GalWeight technique which was specifically designed to simultaneously maximize the number of {\it{bona fide}} cluster members while minimizing the number of contaminating interlopers (Abdullah+18). In Abdullah+18, we applied our GalWeight technique to MDPL2 and Bolshoi N-body simulations and showed that it was $>98\%$ accurate in correctly assigning cluster membership.
The $\mathtt{GalWCat19}$ catalog is at low-redshift for which the effects of cluster evolution and cosmology are minimal. 
Finally, the cluster masses were calculated individually from the dynamics of the member galaxies via the virial theorem (e.g., \citealp{Limber60,Abdullah11}), and corrected for the surface pressure term (e.g., \citealp{The86,Carlberg97}). A huge advantage of our approach relative to mass proxy methods is that it returns an estimate of the total cluster mass (dark matter and baryons) without making any assumptions about the internal complicated physical processes associated with the baryons (gas and galaxies). The publicly available $\mathtt{GalWCat19}$\footnote{\url{https://mohamed-elhashash-94.webself.net/galwcat}}, contains 1800 clusters at redshift $z \le0.2$, which is one of the largest available samples that used a high-quality spectroscopic data set. 

The paper is organized as follows. In \S~\ref{sec:data}, we describe in more detail how we created the $\mathtt{GalWCat19}$ cluster catalog. In \S~\ref{sec:CMF}, we investigate the volume and mass incompleteness of $\mathtt{GalWCat19}$ to obtain a mass-complete local subsample of 756 clusters ($\mathtt{SelFMC}$) used to constrain \om ~and \sig. In \S~\ref{sec:constrain}, we compare our complete sample with theoretical models to constrain the cosmological parameters \om ~and \sig. We investigate how systematics affect the recovered cosmological constraints and compare our results with recent results constrained from some cosmological probes and summarize our conclusions in  \S~\ref{sec:disc}. Throughout the paper we adopt $\Lambda$CDM with $\Omega_m=1-\Omega_\Lambda$, and $H_0=100$ $h$ km s$^{-1}$ Mpc$^{-1}$.
\section{The $\mathtt{GalWCat19}$ Cluster Catalog} \label{sec:data}

In this section, we summarize how we created the $\mathtt{GalWCat19}$ cluster catalog. Full details may be found in Abdullah+20. Using photometric and spectroscopic databases from SDSS- DR13, we extracted data for 704,200 galaxies. These galaxies satisfied the following set of criteria: spectroscopic detection, photometric and spectroscopic classification as galaxy (by the automatic pipeline), spectroscopic redshift between 0.001 and 0.2 (with a redshift completeness $> 0.7$, \citealp{Yang07,Tempel14}), r-band magnitude (reddening-corrected) $< 18$, and the flag SpecObj.zWarning is zero indicating a well-measured redshift. 
 
Galaxy clusters were identified by the well-known Finger-of-God effect (\citealp{Jackson72,Kaiser87,Abdullah13}). The Finger-of-God effect causes a distortion of line-of-sight velocities of galaxies in the redshift-phase space due to the cluster potential well. As described in Abdullah+20, we calculated the membership of each cluster as follows. We firstly calculated the galaxy number density within a cylinder of radius 0.5 \h, and height $3000$ \ks  ~centered on a galaxy, i. Secondly, we sorted all galaxies descending from highest to lowest number densities with the condition that the cylinder has at least 8 galaxies. Thirdly, starting with the galaxy with the highest number density, we applied the binary tree algorithm (e.g., \citealp{Serra11}) to accurately determine a cluster center ($\alpha_{c}$, $\delta_{c}, z_c$) and a phase-space diagram. Fourthly, we applied the GalWeight technique (Abdullah+18) to galaxies in the phase-space diagram out to a maximum projected radius of 10 \h ~and a maximum line-of-sight velocity range of $\pm3000$~\ks~to identify cluster membership. In Abdullah+18, we showed that the  cumulative completeness of the FOG algorithm which we tested using the Bolshoi simulation \citet{Klypin16} was approximately 100\% for clusters with masses $M_{200} > 2 \times 10^{14}~h^{-1}M_{\odot}$, and $\sim 85\%$ for clusters with masses $M_{200} > 0.4 \times 10^{14}~h^{-1}M_{\odot}$.

The virial mass of each cluster was estimated by applying the virial theorem to the cluster members, under the assumption that the mass distribution follows the galaxy distribution (e.g., \citealp{Giuricin82,Merritt88}). The estimated mass was corrected for the surface pressure term which, otherwise, would overestimate the fiducial cluster mass (e.g., \citealp{The86,Binney87,Carlberg97}). The cluster virial mass was calculated at the viral radius within which the cluster is in hydrostatic equilibrium. The virial radius is approximately equal to the radius at which the density $\rho=\Delta_{200}\rho_c$, where $\rho_c$ is the critical density of the universe and $\Delta_{200} = 200$ (e.g., \citealp{Carlberg97,Klypin16}). Abdullah+20 showed that the cluster mass estimates returned by the virial theorem after utilizing the GalWeight technique (Abdullah+18) performed very well in comparison to most of other mass estimation techniques described in \citealp{Old15}. In particular, our procedure was applied to two mock catalogs (HOD2 and SAM2) recalled from \citet{Old15}. We found that the root mean square differences of the recovered mass by GalWeight relative to the fiducial cluster mass were 0.24 and 0.32 for HOD2 and SAM2, respectively. Also, the intrinsic scatter in the recovered mass was $\sim0.23$ dex for both catalogs. Moreover, the uncertainty of the virial mass estimator is calculated using the limiting fractional uncertainty $\pi^{-1}\sqrt{2\ln{N}/N}$ \citep{Bahcall81}.

The scatter and bias in the recovered mass using the virial mass estimator are caused by some factors including: (i) the assumption of hydrostatic equilibrium, projection effect, and possible velocity anisotropies in galaxy orbits, and the assumption that halo mass follows light (or stellar mass); (ii) the presence of substructure and/or nearby structure such as cluster, supercluster, to which the cluster belongs, or filament (e.g., \citealp{Merritt88,Fadda96}); (iii) the presence of interlopers in the cluster frame due to the triple-value problem, for which there are some foreground and background interlopers that appear to be part of the cluster body because of the distortion of phase space \citep{Tonry81,Abdullah13}; and (iv) the identification of cluster center (e.g., \citealp{Girardi98a,Zhang19}).

\begin{figure*}\hspace*{0.25cm}
\includegraphics[width=22 cm]{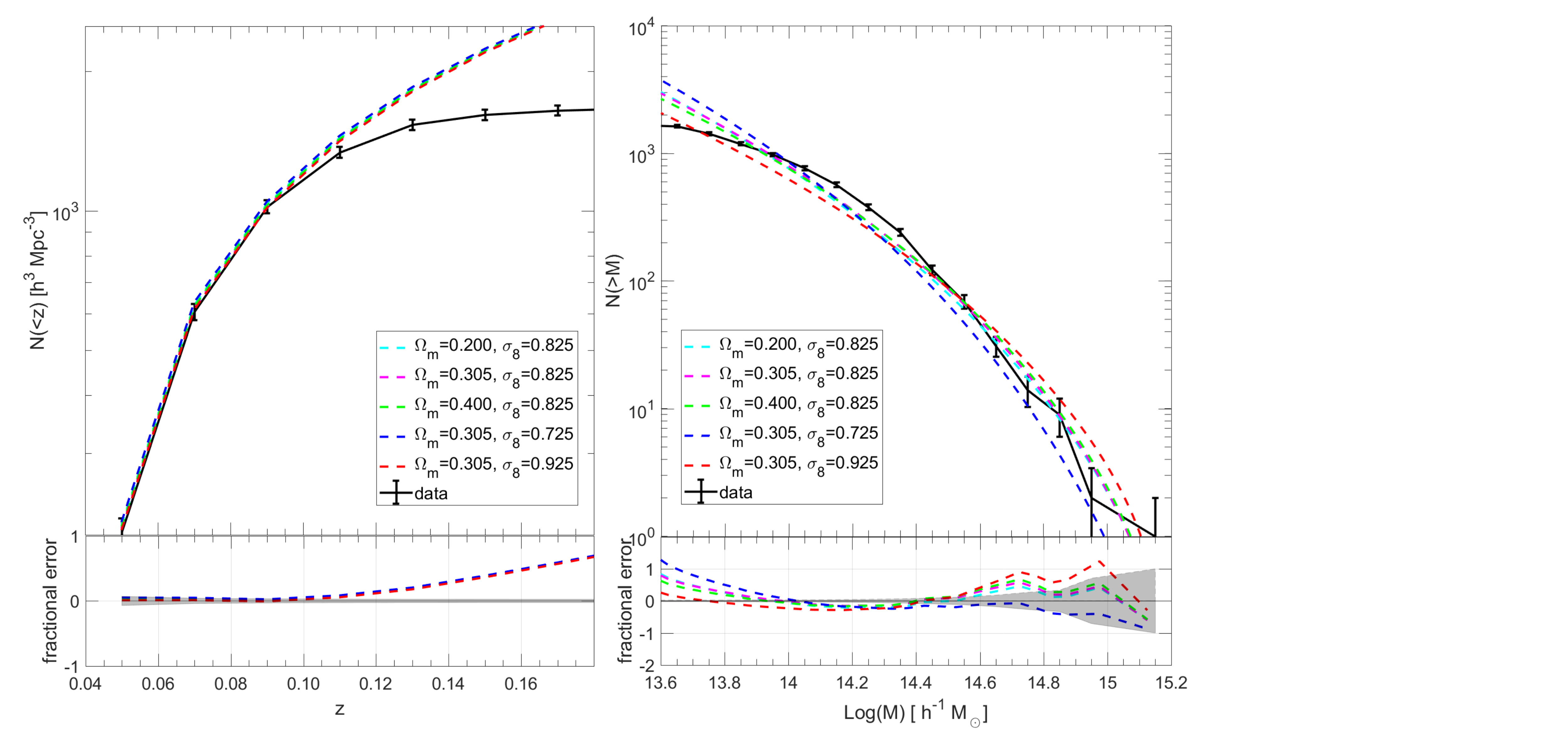} \vspace{-0.5cm}
\caption{$\mathtt{GalWCat19}$ completeness. {\bf{Left}}: The black line shows the integrated abundance of clusters as a function of redshift for the $\mathtt{GalWCat19}$ catalog. The dashed color lines present the expectation of complete samples estimated by Tinker08 for five different cosmologies as shown in the legend. {\bf{Right}}: The black line shows the integrated abundance of clusters as a function of cluster mass. The dashed color lines present the expectation of complete samples estimated by Tinker08 for five different cosmologies as shown in the legend. 
The fractional error $(N(<z)_{obs}-N(<z)_{model})/N(<z)_{model}$ is shown in the lower panels. The gray shaded areas represent the expected Poisson noise.}
\label{fig:Fig01}
\end{figure*}

The 1800 $\mathtt{GalWCat19}$ clusters range in redshift between $0.01 - 0.2$ and in mass between $(0.4 - 14) \times 10^{14}h^{-1}M_{\odot}$. The $\mathtt{GalWCat19}$ catalog contains a large number of cluster parameters including sky position, redshift, membership, velocity dispersion, and mass at overdensities $\Delta = 500, 200, 100, 5.5$. The 34,471 member galaxies were identified within the radius at which the density is 200 times the critical density of the universe. The galaxy catalog provided the coordinates of each galaxy and the ID of the cluster that the galaxy belongs to. The catalogs was publicly available at the following website \url{https://mohamed-elhashash-94.webself.net/galwcat/}.
\section{ Cluster mass function}\label{sec:CMF}

The $\mathtt{GalWCat19}$ catalog is not complete in either volume or mass. In \S~\ref{sec:SF}, we analyze $\mathtt{GalWCat19}$ to develop an appropriate selection function of our sample which is used to correct for the volume incompleteness.  Also, in \S~\ref{sec:PCMF}, we compute the CMF derived from $\mathtt{GalWCat19}$ and compare it with the CMF calculated from the MDPL2 \footnote{\url{https://www.cosmosim.org/cms/simulations/mdpl2/}} simulation (described in the next paragraph) to obtain a mass-complete subsample ($\mathtt{SelGMC}$) used to constrain the cosmological parameters \om ~and \sig.

The MDPL2 is an N-body simulation of $3840^3$ particles in a box of comoving length 1 $h^{-1}$ Gpc, mass resolution of $1.51 \times 10^9$ $h^{-1}$ M$_{\odot}$, and gravitational softening length of 5 $h^{-1}$ kpc (physical) at low redshifts from the suite of MultiDark simulations (see Table 1 in \citealp{Klypin16}). It was run using the L-GADGET-2 code, a version of the publicly available cosmological code GADGET-2  \citep{Springel05}. It assumes a flat $\Lambda$CDM cosmology, with cosmological parameters $\Omega_\Lambda$ = 0.693, $\Omega_m$ = 0.307, $\Omega_b$ = 0.048, $n$ = 0.967, $\sigma_8$ = 0.823, and $h$ = 0.678 \citep{Planck14}. Haloes and subhaloes have been identified with ROCKSTAR \citep{Behroozi13a} and merger trees constructed with CONSISTENT TREES \citep{Behroozi13b}. The catalogs are split into 126 snapshots between redshifts $z = 17$ and $z = 0$. We downloaded the snapshot (hlist\_0.91520.list\footnote{\url{https://www.cosmosim.org/data/catalogs/NewMD_3840_Planck1/ROCKSTAR/trees/hlists/}}) with $z\sim 0.09$ which is consistent with the mean redshift of $\mathtt{GalWCat19}$ sample.

\subsection{$\mathtt{GalWCat19}$ Completeness} \label{sec:SF} 
The $\mathtt{GalWCat19}$ catalog is incomplete in the distribution of clusters with respect to comoving distance (redshift), and in the distribution of clusters with respect to mass. In this section, we discuss such incompleteness and how to make corrections.

The completeness in comoving volume (redshift) of the $\mathtt{GalWCat19}$ catalog can be investigated by calculating the abundance of clusters predicted by a theoretical model and comparing it with the abundance of $\mathtt{GalWCat19}$ clusters. We adopt the functional form of \citet{Tinker08} (hereafter Tinker08) to calculate the halo mass function (HMF\footnote{We use CMF for mass functions derived from observations and HMF for mass functions computed by theoretical models}, see \S ~\ref{sec:Pred} for more details) and consequently the predicted abundance of clusters.

The integrated abundance of clusters as a function of redshift for the $\mathtt{GalWCat19}$ sample, $N(<z)$, is presented in the upper left panel of Figure \ref{fig:Fig01}. Note that $N(<z)$ is calculated for the clusters with redshift $z\geq 0.04$ to remove the effect of nearby regions where the cosmic variance has a large effect due to the small volume. The plot shows that the catalog is matched with the prediction of Tinker08 for $z\lesssim 0.09$. 
Also, the fractional error of $N(<z)$ relative to the expectation of Tinker08, $(N(<z)_{obs}-N(<z)_{model})/N(<z)_{model}$, for each model and the expected Poisson noise (gray shaded area) are presented in the lower left panel. 
The plot shows that the scatter relative to each model is nearly constant (around zero) for $z\lesssim0.09$ before it blows up after this redshift limit. This indicates that $\mathtt{GalWCat19}$ is approximately complete in volume for $z\lesssim 0.09$ (or equivalently comoving distance of $D\lesssim 265$ \h ~for the $\Lambda$CMD universe with \om = 0.3). 
We call this volume-complete subsample as $\mathtt{NoSelFVC}$.

Similarly, the integrated abundance of clusters as a function of cluster mass, $N(>M)$, is presented in the upper right panel of Figure \ref{fig:Fig01} in comparison to five Tinker08 models and the scatter is presented in the lower right panel. The plot shows that the data is matched with the models of \om = [0.20, 0.305, 0.40] with \sig = 0.825 better than the models of \om = 0.305 and \sig = [0.725, 0.925]. Even though it is not an easy task to specifically determine the mass threshold at which the catalog is complete, the three matched models indicate that $\mathtt{GalWCat19}$ is approximately complete for $\log(M)\gtrsim13.9$ \hm. We discuss the systematics of adapting this mass threshold on our analysis in \S ~\ref{sec:sys}. The large scatter at the high mass end is due to the small number of massive clusters, while the large scatter at the low mass end comes from the incompleteness of $\mathtt{GalWCat19}$. 

\begin{figure} \hspace*{-.75cm}
\includegraphics[width=23 cm]{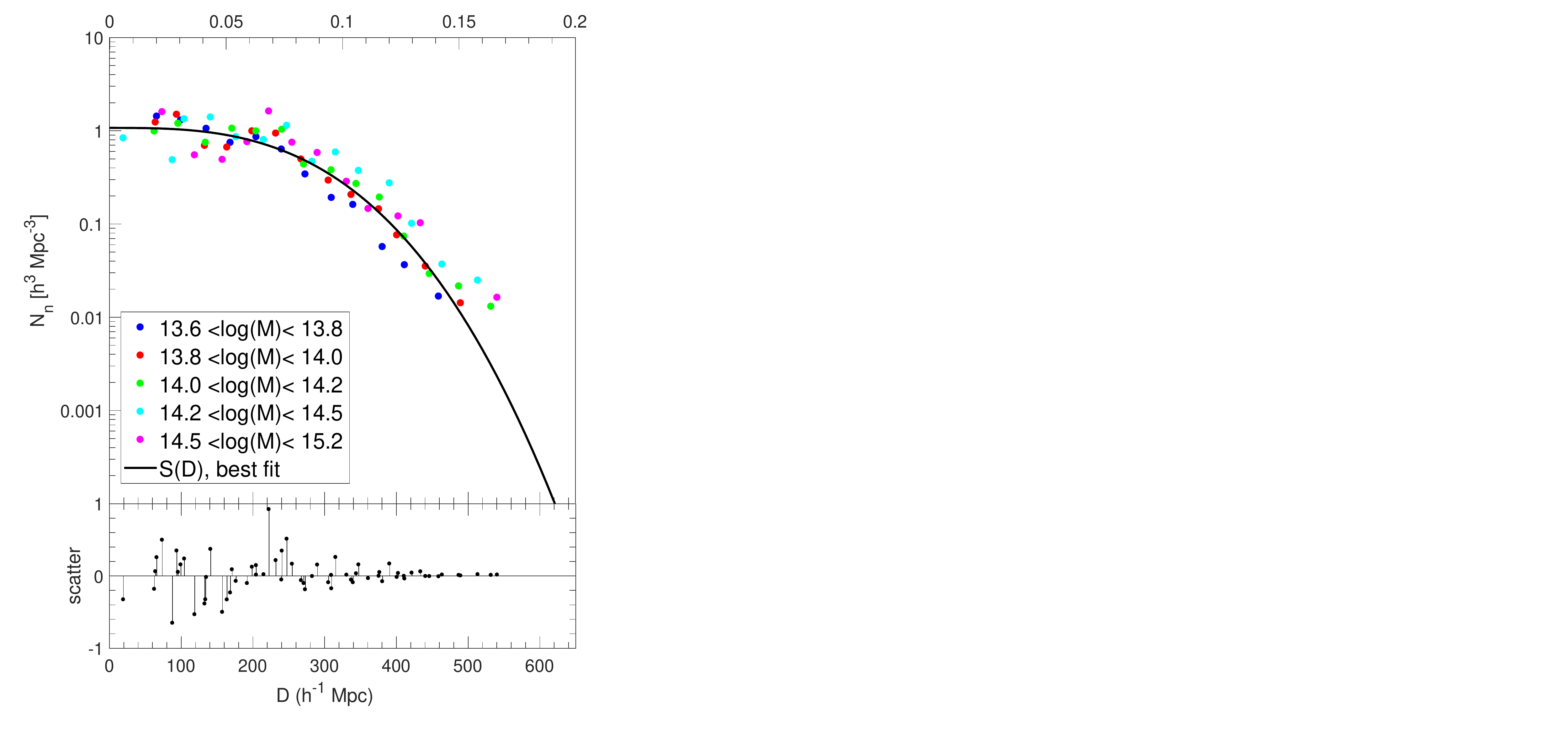} \vspace{-1.cm}
\caption{Selection function of $\mathtt{GalWCat19}$ cluster sample. Colored points show the normalized number density of the five mass bins described in Figure \ref{fig:Fig01}. The black line shows an exponential form describing the selection function $\mathcal{S}(D)$ which is fitted with the data. The scatter of data relative to the exponential form is presented in the lower panel.}
\label{fig:Fig02}
\end{figure}

In order to correct for the incompleteness in volume of $\mathtt{GalWCat19}$ each cluster should be weighed by $\mathcal{S}(D)$, where $\mathcal{S}$ is the selection function at a distance $D$. Figure \ref{fig:Fig02} introduces the normalized number density $\mathcal{N}_n(D)$, defined as the cluster number density normalized by the average number density calculated for clusters within comoving distance $D<265$ \h, for all clusters and for five mass bins as described in Table \ref{tab:ND}. The distribution of points in Figure \ref{fig:Fig02} can be described by an exponential function that represents the selection function $\mathcal{S}(D)$.  It has the form 

\begin{equation} \label{eq:SF}
\mathcal{S}(D) = a \exp\left[{-\left(\frac{D}{b}\right)^\gamma}\right]
\end{equation}

\noindent The parameters $a$, $b$ and $\gamma$ are determined by applying the chi-squared algorithm using the Curve Fitting MatLab Toolbox. The best fit values of these parameters are, $a = 1.07\pm0.12$, $b = 293.4\pm20.7$ \h ~and $\gamma = 2.97\pm0.90$ with root mean square error of 0.15. 
Note that the normalization $a$ is greater than unity because of the scatter and the effect of the cosmic variance. But, we apply the selection function with the condition that $\mathcal{S}(D) \leq 1$.

We should be cautious in using $\mathcal{S}(D)$ at large distances. This is because $S(D \gtrsim 500)$ \h ~drops to $\gtrsim 0.01$ as demonstrated in Figure \ref{fig:Fig02} which means that a distant cluster would be weighted as at least 100 times as a nearby cluster. This will overestimate or overcorrect the number of clusters at large distances, and consequently the estimated CMF will be noisy. Thus, in order to avoid the overcorrection and the noisiness of CMF we restrict our sample to a maximum comoving distance of $D\leq365$ (or $z\leq0.125$) for which $\mathcal{S}(D)\lesssim0.2$. 

It is well-known that the cluster number density of a given mass decreases with redshift for a 100\% complete sample because of the HMF evolution effect. Thus, the CMF should be scaled or corrected by an evolution function, $\mathcal{S}_{evo}(D)$. For a sample with a broad range of redshifts, the only way to take the evolution into account is to calculate this function. However, the disadvantage of this approach is that the correction is model dependent: the measured HMF (i.e., CMF) is a convolution of the true HMF and theoretical estimate of $\mathcal{S}_{evo}(D)$. However, for a sample with a narrow range of redshifts (as in our case) we show in appendix \ref{app:evo} that the evolution effect is less than 3\% for clusters in the redshift range of $0.045 \leq z \leq 0.125$. In appendix \ref{app:AA}, we discuss the effect of adopting this redshift interval on our results.


\begin{table} \centering
\caption{The cluster average number density for different mass bins.}
\label{tab:ND}
\begin{tabular}{cccc} 
\hline
Mass & number of& average & color \\
bin & clusters & number density & \\
$[$\hm $]$ & & [$10^{-5}~h^{3}$ Mpc$^{-3}$]&\\
\hline
13.6 - 15.2& 1800 & 5.6& black\\ \hline
13.6 - 13.8 & 527 & 2.2 & blue   \\
13.8 - 14.0 & 461  & 1.5 & green\\
14.0 - 14.2  & 411 & 1.0 & red    \\
14.2 - 14.5  & 326 & 0.7 & cyan \\
14.5 - 15.2  & 75   &  0.2 & magenta\\
\hline
\end{tabular}
\begin{tablenotes}
\item
\noindent
Columns: (1) the mass bin in units of $\log \mbox{M}$ [\hm]; (2) the number of clusters in each mass bin; (3) the average number density calculated for clusters within comoving distance $D<265$ \h ~in each mass bin; (4) the color of number density profile as shown in the right panel of Figure \ref{fig:Fig01}.
\end{tablenotes}
\end{table} 

\subsection{Estimating the Mass Function} \label{sec:PCMF}
In this section, we compute the CMF, d$n(M)/$d$\log(M)$, and its corresponding cumulative mass function, $n(>M)$, which are estimated for a  $\Lambda$CDM cosmology with \om$=0.3$ and $\Omega_\Lambda=0.7$. The CMF is defined as the number density of clusters per logarithmic cluster mass interval. Also, the cumulative CMF is defined as the number density of clusters more massive than a given mass $M$.

Mathematically, the CMF, weighted by the selection function $\mathcal{S}$, is given by

\begin{equation} \label{eq:MF}
\frac{dn(M)}{d\log{M}} = \frac{1}{d\log{M}}\sum_i\frac{1}{V}\frac{1}{\mathcal{S}(D_i)}
\end{equation}
\noindent where $D_i$ is the comoving distance of a cluster i, and $V$ is the comoving volume which is given by

\begin{equation} \label{eq:VV}
V = \frac{4 \pi}{3} \frac{\Omega_{survey}}{\Omega_{sky}}(D_2^3-D_1^3)
\end{equation}

\noindent where $\Omega_{sky} = 41,253$ deg$^2$ is the area of the sky, $\Omega_{sur}\simeq 11,000$ deg$^2$ is the area covered by $\mathtt{GalWCat19}$, and $D_1$ and $D_2$ are the minimum and maximum comoving distances of the cluster sample.

\begin{figure} \hspace*{-0.0cm}
\includegraphics[width=22 cm]{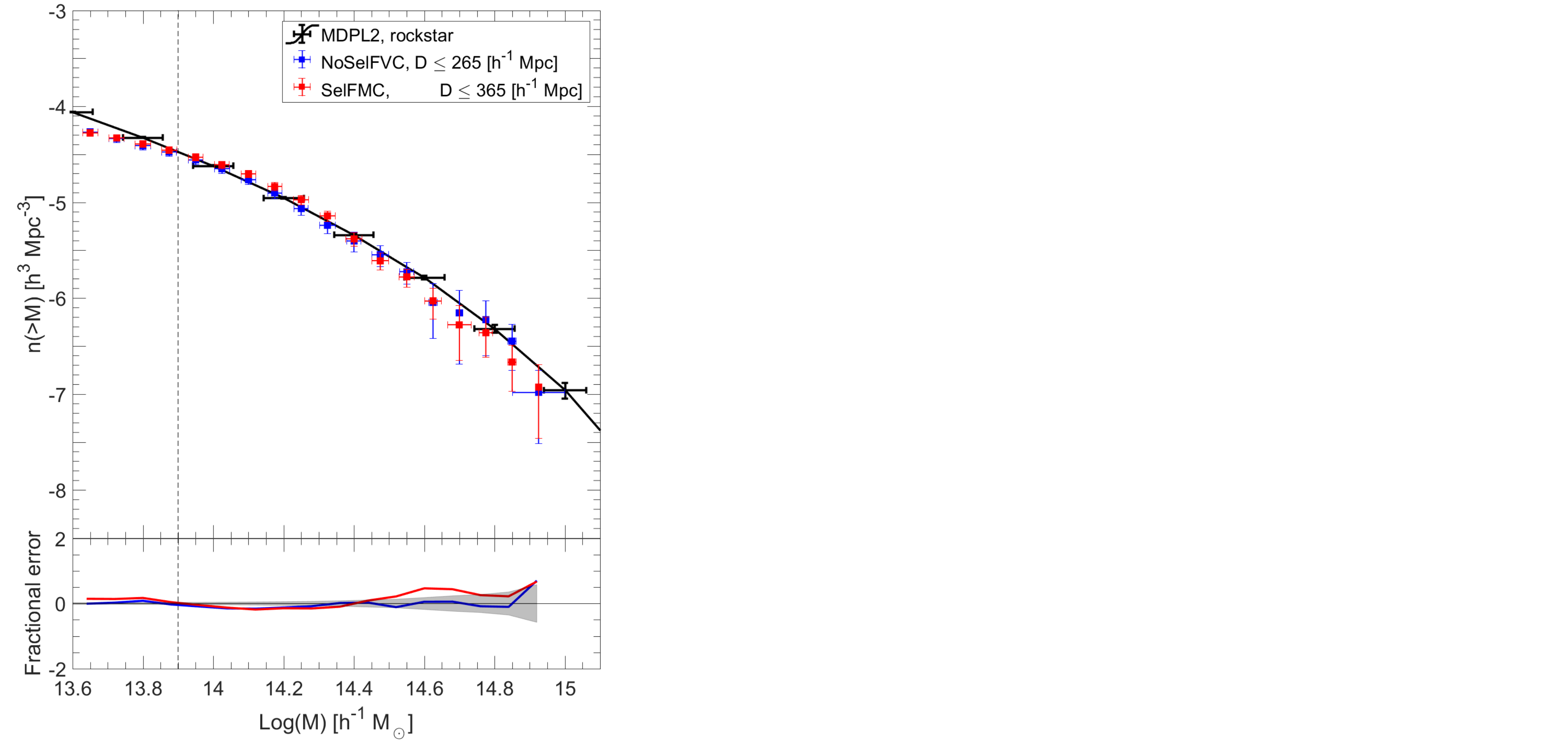} \vspace{-0.25cm}
\caption{The cumulative CMF derived from the $\mathtt{GalWCat19}$ cluster sample. The black line shows the mass function computed from the MDPL2 simulation (for the snapshot hlist\_0.91520.list at $z \sim 0.9$ or $D \sim 260$) \citep{Klypin16}. The blue points present the CMF for the volume-complete subsample with $D \leq 265$ \h ~($z \sim 0.09$) without the correction of $\mathcal{S}(D)$ ($\mathtt{NoSelFVC}$). The red points show the CMF corrected by $\mathcal{S}(D)$ for $D \leq 365$ \h ($z \sim 0.125$, $\mathtt{SelFMC}$). The vertical dashed line shows the low-mass limit ($\log(M) = 13.9$ \hm) used to constrain \om ~and \sig. The error bars on the vertical axis are calculated by Poisson statistics. 
The fractional errors between the CMF of MDPL2 and both $\mathtt{NoSelFVC}$ and $\mathtt{SelFMC}$ are shown in the lower panels. The gray shaded areas represent the expected Poisson noise.}
\label{fig:Fig03}
\end{figure}

Figure \ref{fig:Fig03} introduces the cumulative CMF computed from $\mathtt{GalWCat19}$. The black line is the CMF computed from the MDPL2 simulation (for the snapshot hlist\_0.91520.list at $z \sim 0.09$ or $D \sim 265$, \citealp{Klypin16}). The blue points introduces the CMF for $\mathtt{NoSelFVC}$ without the correction of $\mathcal{S}(D)$, since this sample is already complete in volume (see, \S~\ref{sec:SF} and Figure \ref{fig:Fig01}). The red points represents our CMF corrected by $\mathcal{S}(D)$ for $D \leq 365$ \h ~($z \sim 0.125$). 
Comparing the CMF estimated by the $\mathtt{NoSelFVC}$ subsample with that derived from the MDPL2 simulation indicates that the sample is  approximately complete in mass for $\log(M)\gtrsim13.9$ \hm, while it drops lower than the CMF of MDPL2 at low-mass end. Also, our CMF, corrected by $\mathcal{S}(D\leq 365)$, is in good agreement with the CMF derived from $\mathtt{NoSelFVC}$ with a scatter of 0.026 dex. 
The mass completeness of $\mathtt{GalWCat19}$ is discussed in \S ~\ref{sec:SF} and Figure \ref{fig:Fig01}. 
In appendix \ref{app:AA}, we show that the results of deriving the cosmological parameters from  $\mathtt{NoSelFVC}$ is consistent with that derived from $\mathtt{SelFMC}$. This indicates that weighting each cluster in our sample by $S(D\leq365)$ introduced in \S~\ref{sec:SF} and Equation \ref{eq:SF} is sufficient to correct for the volume incompleteness of $\mathtt{GalWCat19}$. 

Therefore, our final subsample, corrected by $S(D)$ is restricted by $\log(M)\geq13.9$ \hm ~and $0.045 \leq z \leq 0.125$. The number of clusters of this subsample is 756, which represents $\sim 42\%$ of the $\mathtt{GalWCat19}$ sample. We use this subsample to constrain \om ~and \sig ~and call it as fiducial $\mathtt{SelFMC}$ sample.

\section{Implications for Cosmological Models}  \label{sec:constrain}
In \S~\ref{sec:Pred}, we discuss the prediction of HMF from the theoretical framework. In \S~\ref{sec:constrain} we derive the constrains on the cosmological parameters \om ~and \sig, and discuss the degeneracy between these two parameters.

\subsection{Prediction of Halo Mass Function} \label{sec:Pred}

\begin{figure} \hspace*{-0.25cm}
\includegraphics[width=18.5cm]{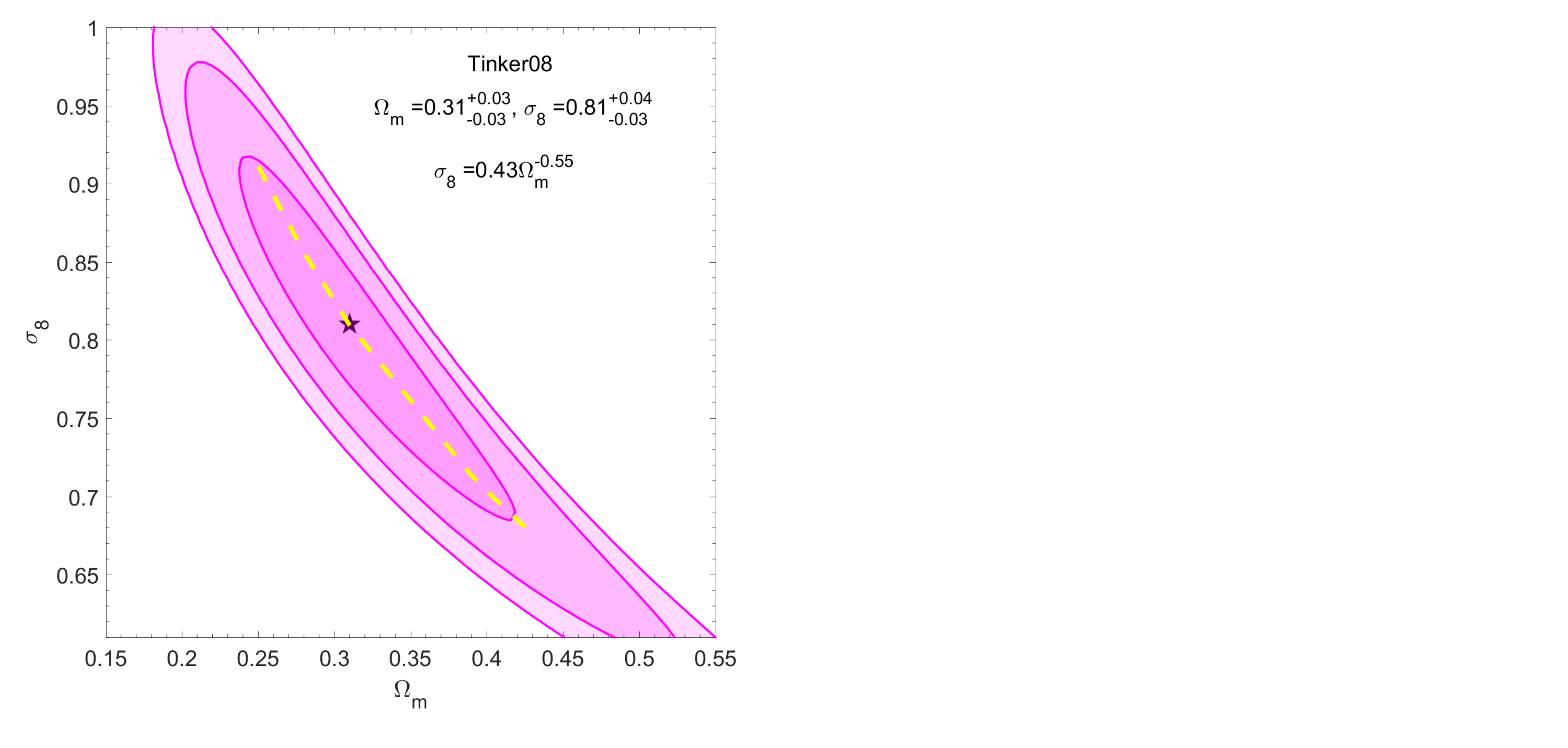} \vspace{-0.75cm}
\caption{Likelihood contour map of $\chi^2$ in \sig-\om ~plane derived from the $\mathtt{SelFMC}$ cluster catalog. The black star represents the best-fit point for \om ~and \sig~ which minimizes $\chi^2$ value. Ellipses show 1$\sigma$, 2$\sigma$, and 3$\sigma$ confidence levels, respectively. The dashed yellow line represents the best-fit \sig-\om ~relation as shown in the legend.}
\label{fig:Fig04}
\end{figure}

The number of dark matter halos per unit mass per unit comoving volume of the universe, HMF, is given by 

\begin{equation}
\label{eq:hmf}
  \frac{dn}{d\ln M} =  f(\sigma) \frac{\rho_0}{M} \left|\frac{d\ln\sigma}{d\ln M}\right|;
\end{equation}

\noindent here $\rho_0$ is the mean density of the universe, $\sigma$ is the rms mass variance on a scale of radius $R$ that contains mass $M = 4 \pi \rho_0 R^3/3$ , and $f(\sigma)$ represents the functional form that defines a particular HMF fit.
 
Assuming a Gaussian distribution of mass fluctuation, \citet{Press74} used a linear theory to derive the first theoretical model (hereafter PS) of HMF. While fairly successful in matching the results of N-body simulations, the PS formalism tends to predict too many low-mass clusters and too few high-mass clusters. More recently proposed theoretical models provide better approximations to the output from N-body simulations (e.g., \citealp{Sheth01,Jenkins01,Warren06,Tinker10b,Bhattacharya11,Behroozi13a}). 

\begin{figure*} \hspace*{0.5cm}
\includegraphics[width=21 cm]{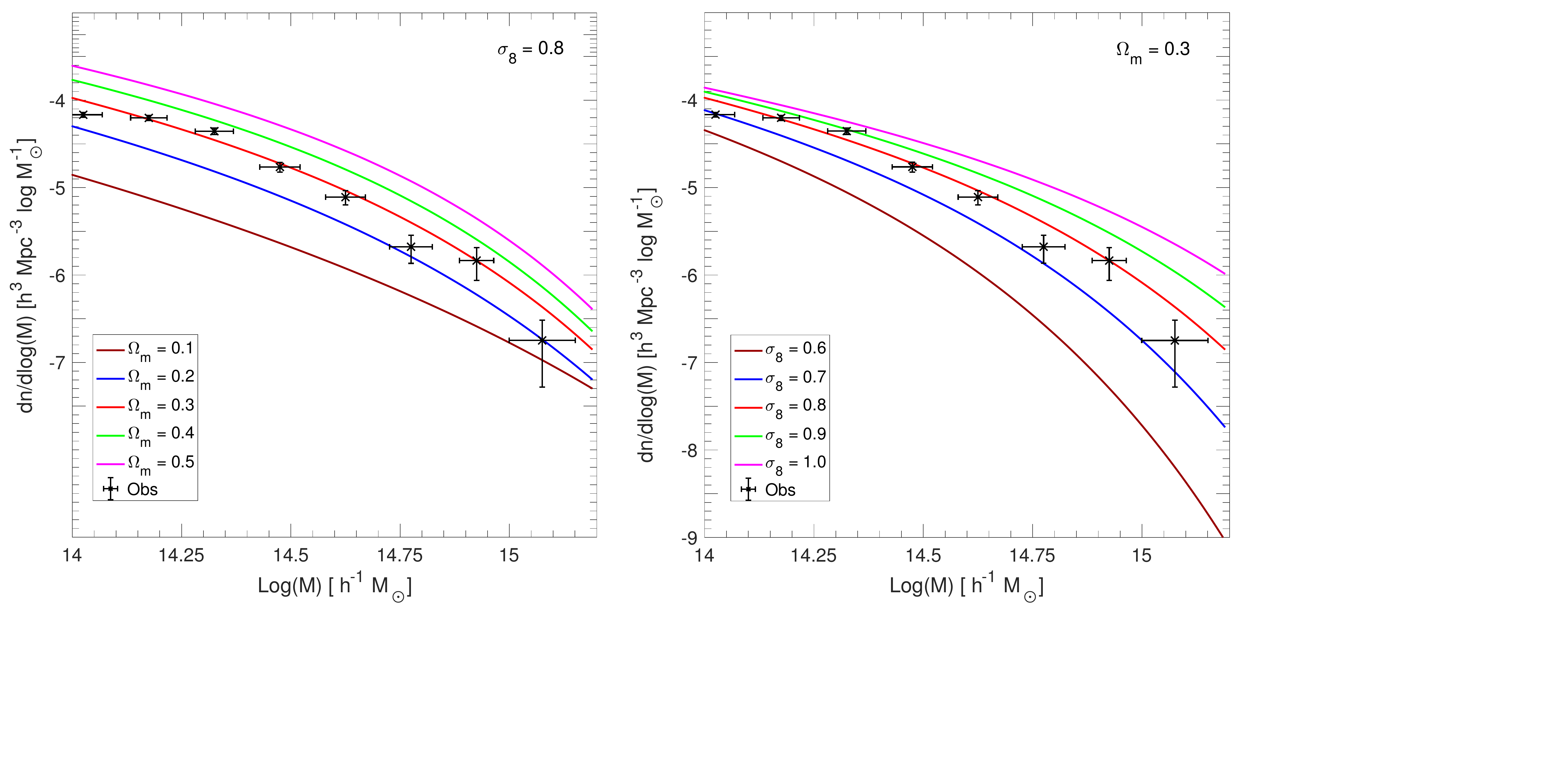} \vspace{-2.35cm}
\caption{Effect of varying \om ~and \sig ~on the HMF. The left panel shows the HMF calculated from Tinker08 for five different values of \om = [0.1 0.2 0.3 0.4 0.5] while fixing \sig= 0.3 (solid colored lines as shown in the legend). The right panel shows the HMF calculated from Tinker08 for five different values of \sig = [0.6 0.7 0.8 0.9 1.0] while fixing \om = 0.3 (solid colored lines as shown in the legend). Our derived CMF corrected by $\mathcal{S}(D)$ for $D \leq 365$ \h ($z \sim 0.125$) is shown by black points.}
\label{fig:Fig05}
\end{figure*}

In this paper, we adopt the functional form proposed by \citet{Tinker08} (hereafter Tinker08) as our form of the HMF. This approach assumes universality of the HMF across the cosmological parameter space considered in this work, and uses a fitting function that was calibrated against N-body simulations. The Tinker08 model is formally accurate to better than 5\% for the cosmologies close to the $\Lambda$CDM cosmology and for the mass and redshift range of interest in our study (e.g., \citealp{Vikhlinin09}). Although the formula has been calibrated using dissipationless N-body simulations (i.e., without the effect of baryons), hydrodynamic simulations suggest that these have negligible impact for clusters with masses as high as those considered here (e.g., \citealp{Rudd08,Velliscig14,Bocquet16}). Finally, note that the Tinker08 model is defined in spherical apertures enclosing overdensities similar to the mass we derive for the $\mathtt{GalWCat19}$ observed clusters.

\begin{equation}\label{eq:T08}
f(\sigma,z) = A\left[\left(\frac{\sigma}{b}\right)^{-a}+1\right] \exp{(-c/\sigma^2)}
\end{equation}  

\noindent where $A = 0.186\left(1+z\right)^{-0.14}$, $a = 1.47\left(1+z\right)^{-0.06}$, $b = 2.57\left(1+z\right)^{-\alpha}$, $c =1.19$, and $\ln{\alpha}(\Delta_{vir}) = \left[75 / \left(\ln{(\Delta_{vir}/75)}\right)\right]^{1.2}$, and $\sigma^2$ is the mass variance defined as

\begin{equation}\label{eq:s2}
\sigma^2(M,z) =\frac{g(z)}{2\pi} \int P(k)W^2(kR)k^2dk
\end{equation}

\noindent $P(k)$ is the current linear matter power spectrum (at $z=0$) as a function of wavenumber $k$, $W(kR) = 3\left[\sin(kR) - kR\cos(kR)\right])/(kR)^3$ is the Fourier transform of the real-space top-hat window function of radius R, and $g(z)=\sigma_8(z)/\sigma_8(0)$ is the growth factor of linear perturbations at scales of 8\h, normalized to unity at $z = 0$.

The current linear power spectrum $P(k)$ is defined as $P(k) = B k^n T^2(k)$, where $T(k)$ is the transfer function, $B$ is the normalization constant and $n$ is the spectral index. Usually the normalization $B$ is calculated from the cosmological parameter \sig, (e.g., \citealp{Reiprich02,Murray13b}). The function $k^n$ imprints the primordial power spectrum during the epoch of inflation. The transfer function $T(k)$ quantifies how this primordial form is  evolved with time to the current linear power spectrum on different scales. The transfer function $T(k)$ is calculated using the public \textit{Code for Anisotropies in the Microwave Background} (CAMB\footnote{\url{https://camb.info/}}, \citealp{Lewis00}). The quantities \om ~and \sig ~are the main cosmological parameters that define the HMF. The other parameters do not strongly affect the HMF and thus we fix them during the calculation of the HMF as described below (e.g., \citealp{Reiprich02,Bahcall03,Wen10}).

\subsection{Constraining \om~and \sig} \label{sec:const}
The HMF is calculated using the publicly available \verb|HMFcalc| \footnote{\url{http://hmf.icrar.org/}} code \citep{Murray13b}. The code provides about 20 fitting functions that can be used to calculate the HMF. In this paper, in order to constrain \om ~and \sig, we use Tinker08 (Equation \ref{eq:T08}) as discussed above. We calculate the HMF by allowing \om ~to range between [0.1, 0.6]  and \sig ~between [0.6, 1.2], both in steps of 0.005. We keep the following cosmological parameters fixed: the CMB temperature $T_{cmb}=2.725 K^\circ$, baryonic density $\Omega_b=0.0486$, and spectral index $n = 0.967$ \citep{Planck14}, at redshift $z=0.089$ (the mean redshift of $\mathtt{GalWCat19}$). 


In order to determine the best-fit mass function and constrain \om ~and \sig ~we use a standard $\chi^2$ procedure

\begin{equation}
      \chi^2 = \sum_{i=1}^N\left(\frac{\left[y_{o,i}-y_{m,i}\right]^2}{\sigma_i^2} \right)
\end{equation}

\noindent where the likelihood, $\mathcal{L}(y|\sigma_8,\Omega_m)$, of a data (CMF) given a model (HMF) is 

\begin{equation}
      \mathcal{L}(y|\sigma_8,\Omega_m) \propto \exp{\left(\frac{-\chi^2(y|\sigma_8,\Omega_m)}{2}\right)}
\end{equation}

\noindent $y_o$ and $y_m$ are the data and model cumulative mass functions at a given mass and $\sigma$ is the statistical uncertainty of the data.

Using the fiducial $\mathtt{SelFMC}$ sample of 756 clusters with $\log(M)\geq13.9$ and $0.045 \leq z \leq 0.125$, the best-fit parameters for the minimum value of $\chi^2$ are \om ~$=0.310^{+0.025}_{-0.029}$ and \sig ~$=0.810^{+0.039}_{-0.034}$ for Tinker08 at redshift $z = 0.085$. In \S ~\ref{sec:sys} we discuss the systematics of cluster mass uncertainty, mass threshold, and selection function.

The banana shape in Figure \ref{fig:Fig04} shows the well-known degeneracy between \sig ~and \om. The relationship between \sig ~and \om ~is often expressed as

\begin{equation} \label{eq:sor}
\sigma_8=\alpha~\Omega_m^{\beta}
\end{equation}

\noindent The parameters $\alpha$, $\beta$, and $\delta$ are determined by applying the $\chi^2$ algorithm using the Curve Fitting MatLab. The best fit values of these parameters are $\alpha = 0.425\pm0.006$ and $\beta = - 0.550\pm0.007$ with root mean square error of 0.005 for the Tinker08 model.

We now ask the question - how do \om ~and \sig ~contribute individually to the HMF? In other words, why do cluster abundance studies introduce a degeneracy between \om ~and \sig?  The degeneracy occurs because a low abundance of massive clusters could be caused either by a small amount of matter in the universe (a low value of \om) or small fluctuations in the density field (a low value of \sig). Similarly, a high abundance of massive clusters could be caused either by a large amount of matter in the universe (a high value of  \om) or large fluctuations in the density field (a high value of \sig). Therefore, it is possible to obtain the same abundance of massive clusters by fixing one parameter and varying the other one. Figure \ref{fig:Fig05} introduces two sets of HMFs calculated by Tinker08. The first set is shown on the left panel for five different values of \om ~= [0.1 0.2 0.3 0.4 0.5] while fixing \sig ~= 0.8. The second set is shown on the right panel for five different values of \sig ~= [0.6 0.7 0.8 0.9 1.0] while fixing \om ~= 0.3.  As expected, increasing the matter density of the universe  increases the number of clusters of all masses. But increasing the rms mass fluctuation increases the number of high-mass clusters more dramatically than number the low-mass clusters. In other words, \sig ~is very sensitive to the high-mass end of the HMF. 




\begin{figure*} \hspace*{-.250cm}
\includegraphics[width=23cm]{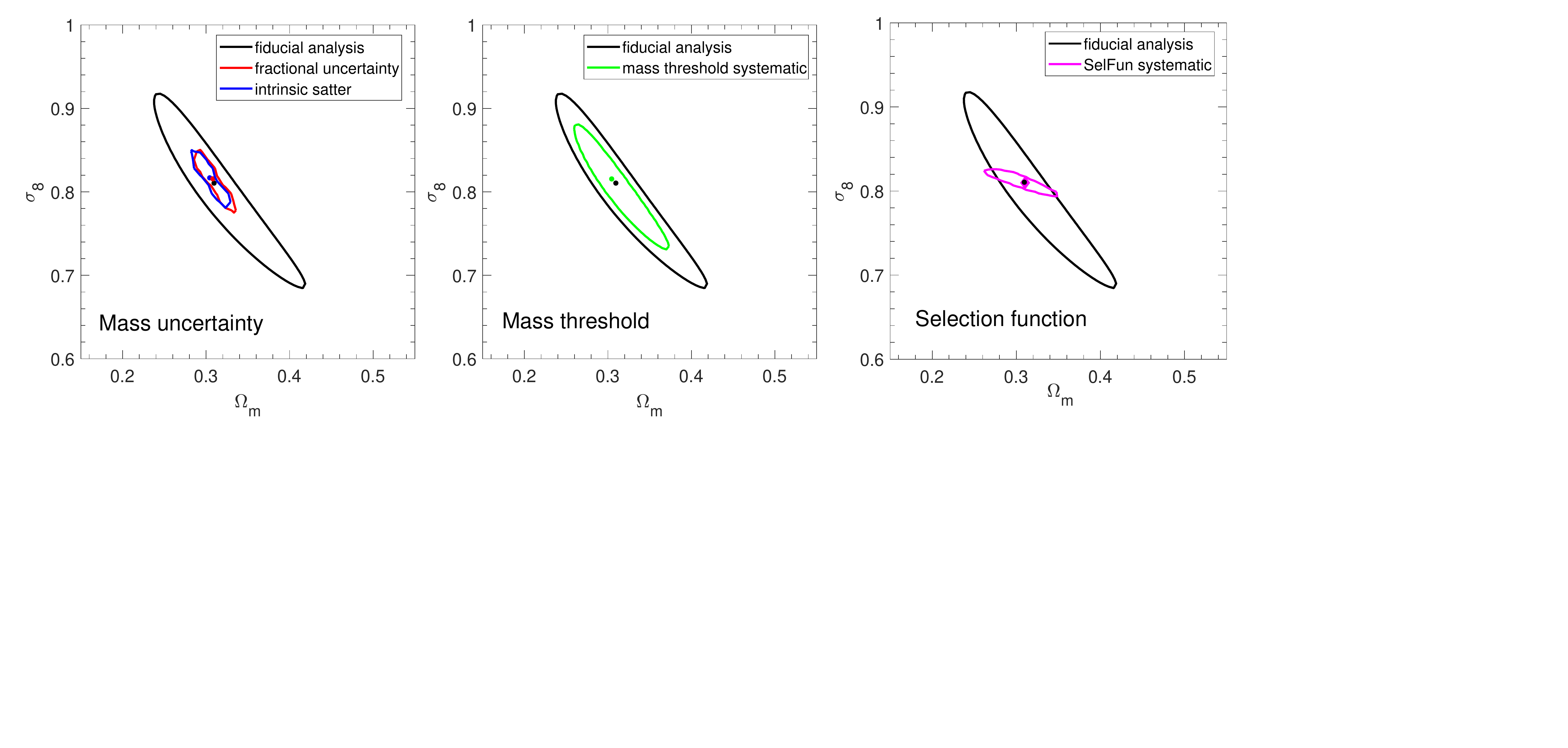} \vspace{-5.25cm}
\caption{Effects of cluster mass uncertainty (left), mass threshold (middle), and selection function (right) on our constraints on \om ~and \sig. {\bf{Left}}: the 68\%  CLs of our fiducial sample (black), fractional mass uncertainty (blue), and intrinsic scatter of 0.23 (red). {\bf{Middle}}: the 68\% CLs (green) for varying mass threshold $\log{M}$ from 13.8 to 14 \hm. {\bf{Right}}: 
the 68\% CLs (magenta) due to systematic of  the selection function.
} 
\label{fig:Fig06}
\end{figure*}

\section{Discussion and Conclusion} \label{sec:disc}

In this section, we investigate how systematics affect the recovered cosmological constraints from our analysis (\S ~\ref{sec:sys}). 
We compare our constraints on the cosmological parameters \om ~and \sig ~with those obtained from cluster abundance studies (\S~\ref{sec:combine}). We also compare our constraints with those obtained from other cosmological probes which we refer to as non-cluster cosmological probes (\S~\ref{sec:combine1}).

\subsection{Systematics} \label{sec:sys}
In constraining \om ~and \sig ~in \S ~\ref{sec:const}, we only account for the statistical uncertainty of the estimated cumulative CMF using the fiducial $\mathtt{SelFMC}$ sample. In this section, we discuss the systematics due to mass uncertainty, mass threshold, and parameterization of the selection function.

\subsubsection{Mass Uncertainty} \label{sec:masserr}
The first uncertainty comes from the difficulty of calculating cluster masses accurately. Generally, masses which are estimated using scaling relations, such as luminosity, richness, temperature, and dispersion velocity-mass relations, introduce large scatter and consequently large systematic uncertainties (e.g., \citealp{Mantz16a,Mulroy19}). Masses which are computed by dynamical estimators are subject to systematic uncertainties (e.g., \citealp{Wojtak07,Rozo10,Old18}). However, using the virial theorem, corrected for the surface pressure term, provides a relatively unbiased estimation of cluster masses (e.g., \citealp{Rines10,Ruel14}), particularly when using a sophisticated interloper rejection technique such as  GalWeight (Abdullah+18). Also, the virial mass estimator calculates the total cluster mass including baryonic (gas and galaxies) and dark matter regardless the internal complex physical processes associated with the baryonic component in clusters. However, the virial mass estimator still introduces scatter in estimating cluster masses (see \S ~\ref{sec:data}). Abdullah+20 showed that the application of the virial mass estimator on two mock catalogs (HOD2 and SAM2) recalled from \citet{Old15} returned intrinsic scatter of $\sim 0.23$ dex in the recovered mass relative to the fiducial cluster mass. Also, the $\mathtt{GalWCat19}$ catalog introduced the fractional uncertainty (see \S ~\ref{sec:data}) of each cluster mass. 

Assuming a normal distribution, we investigate the systematics of the mass uncertainty by generating $\sim 8000$ estimate for each cluster mass using both the fractional uncertainty for each cluster and the intrinsic scatter for the entire sample. 
In other words, we reanalyze $\mathtt{SelFMC}$ $\sim 8000$ times and refit for \om ~and \sig~ for each time. The left panel of Figure \ref{fig:Fig06} introduces the effect of cluster mass uncertainty on the constraints on \om ~and \sig. Using the fractional uncertainty, we obtain \om $= 0.305\pm0.014$ and \sig $= 0.816\pm0.021$, where the red ellipse represents 68\% CL for the disribution of the reestimated 8000  pairs of \om ~and \sig. Using the intrinsic scatter (blue ellipse), we find \om $= 0.309\pm0.014$ and \sig $= 0.815\pm0.022$. Both results indicate that the cluster mass uncertainty (fractional or intrinsic) does not affect our constraints on \om ~and \sig ~using $\mathtt{SelFMC}$.

\subsubsection{Mass Threshold} \label{sec:massthr}
The second systematic uncertainty comes from the difficulty of determining  accurately the mass threshold at which the sample is mass complete. As discussed in \S ~\ref{sec:SF} and Figure \ref{fig:Fig01} the catalog is approximately complete around $\log{M}\gtrsim 13.9$ [\hm]. However, the mass threshold at which the sample is mass-complete is not accurately specified. Therefore, we investigate the effect of varying the mass threshold $\log{M}$ between 13.8 and 14.0 [\hm] ~in steps of 0.05 dex on the recovered cosmological constraints from our analysis.
For each mass threshold we calculate the $\chi^2$ likelihood and then we obtain the joint 68\% CL of all $\chi^2$ distributions as shown in the middle panel of Figure \ref{fig:Fig06}.
The plot shows that the best fit values of \om ~and \sig ~deviate very slightly from the results of the fiducial sample with \om ~$=0.300^{+0.015}_{-0.017}$ and \sig ~$=0.820^{+0.020}_{-0.023}$.

\begin{figure*} \hspace*{0.75cm}
\includegraphics[width=21cm]{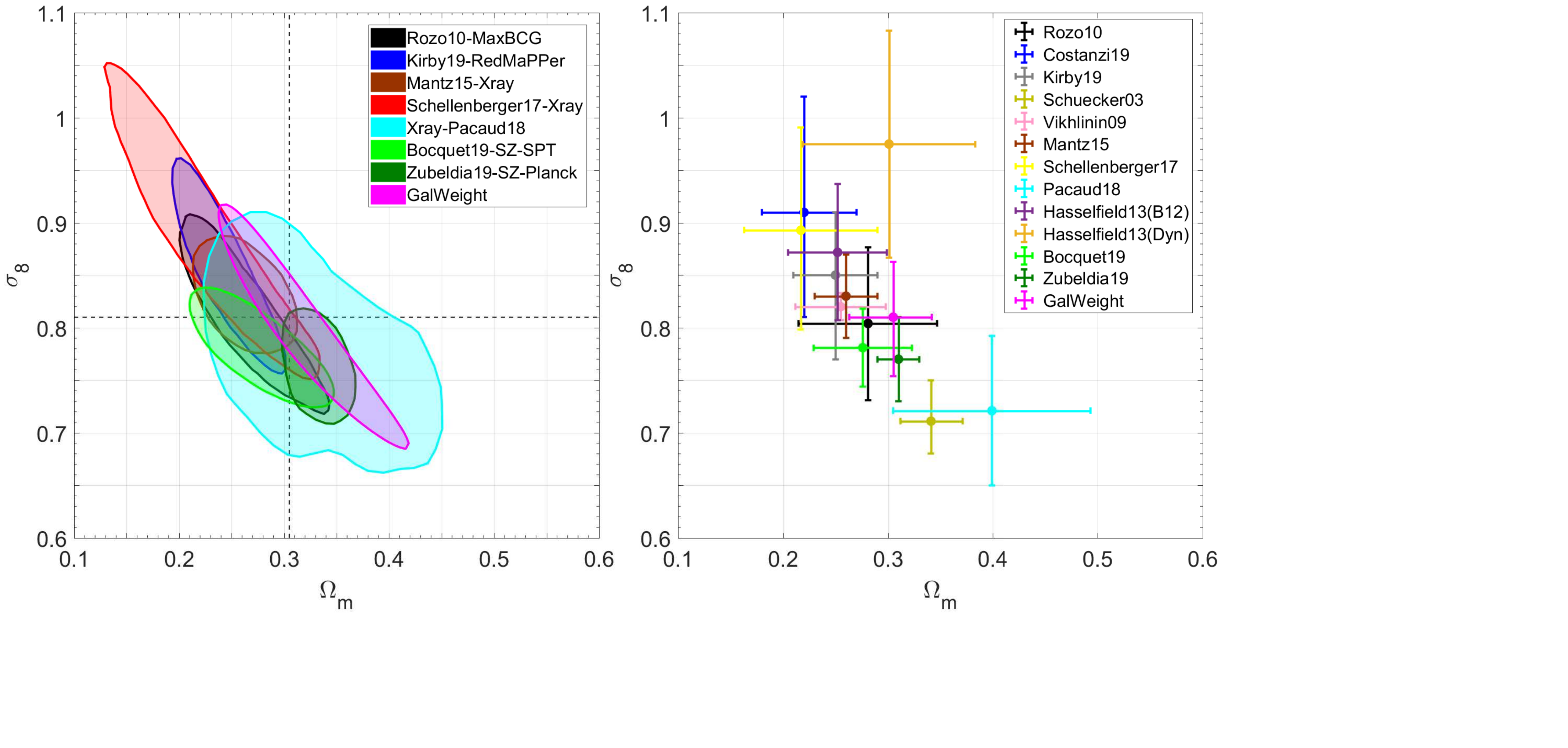} \vspace{-2.15cm}
\caption{Constraints on \om ~and \sig ~obtained from cluster abundance studies (cluster mass function; CMF).  Left: 68$\%$ confidence levels (CLs) derived from $\mathtt{SelFMC}$ (magenta) plus select other optical, X-ray or SZ-detected cluster catalogs as shown in the legend and summarized in the first three sections of Table \ref{tab:Comp}. The two dashed lines show the best-fit values derived in this work. Right: Uncertainties on \om ~and \sig ~for each of the cluster abundance studies listed in Table \ref{tab:Comp} (Note: For clarity, not all studies in Table \ref{tab:Comp} are shown in the left panel). While in agreement with the other cluster abundance studies within 1$\sigma$ uncertainties, the value of \om\ determined from our work is slightly higher and the value of \sig\ slightly lower than most of the other studies. As shown in Fig~\ref{fig:Fig11} and discussed in \S~\ref{sec:combine1}, we note that our values are in better agreement with \om ~and \sig ~obtained from non-cluster determinations as shown in Fig~\ref{fig:Fig11}.
}
\label{fig:Fig10}
\end{figure*}

\begin{table*} \centering
\caption{Comparison of constraints on cosmological parameters \om ~and \sig~derived from Clusters Abundances (CMF) and from Other Cosmological Probes, including cosmic shear, galaxy-galaxy lensing, angular clustering, BAO, supernovae, and CMB}
\label{tab:Comp}
\scriptsize
\begin{tabular}{cccccccc}
\hline
Sample&  Mass estimation & \om & \sig &$S_8^{~(a)}$&$\Delta_{pl}^{~(b)}$&reference\\
\hline
\multicolumn{7}{c}{spectroscopically-selected catalogs --- cluster abundance}\\
\hline
$\mathtt{GalWCat19}$ & virial theorem & $0.305^{+0.037}_{-0.042}$& $0.810^{+0.053}_{-0.056}$&0.817&0.032&This work\\
\hline
\multicolumn{7}{c}{optical photometrically-selected catalogs --- cluster abundance}\\
\hline
MaxBCG     & richness-mass +WL & $0.281^{+0.066}_{-0.066}$& $0.804^{+0.073}_{-0.073}$&0.779&0.108&\citealp{Rozo10}\\
RedMaPPer & richness-mass +WL & $0.220^{+0.050}_{-0.040}$& $0.910^{+0.110}_{-0.100}$&0.778&0.325&\citealp{Costanzi19}\\
RedMaPPer & richness-mass +X-ray & $0.250^{+0.040}_{-0.040}$& $0.850^{+0.06}_{-0.08}$&0.776&0.212&\citealp{Kirby19}\\
\hline
\multicolumn{7}{c}{x-rays catalogs --- cluster abundance}\\
\hline
REFLEX & luminosity-mass & $0.341^{+0.030}_{-0.029}$& $0.711^{+0.039}_{-0.031}$&0.758&0.148&\citealp{Schuecker03}\\
Chandra-ROSAT & luminosity-mass & $0.255^{+0.043}_{-0.043}$& $0.820^{+0.013}_{-0.013}$&0.757&0.191&\citealp{Vikhlinin09}\\
ROSAT (WtG)$^{~(c)}$ & luminosity-mass & $0.260^{+0.030}_{-0.030}$& $0.830^{+0.04}_{-0.04}$&0.773&0.176&\citealp{Mantz15}\\
ROSAT - HIFLUGCS & luminosity-mass & $0.217^{+0.073}_{-0.054}$& $0.893^{+0.098}_{-0.095}$&0.760&0.327&\citealp{Schellenberger17}\\
XMM-XXL & temperature-mass & $0.399^{+0.094}_{-0.094}$& $0.721^{+0.071}_{-0.071}$&0.832&0.289&\citealp{Pacaud18}
\\
\hline
\multicolumn{7}{c}{SZ catalogs --- cluster abundance}\\
\hline
ACT, [BBN+H0+ACTcl(B12)] &SZ-mass & $0.252^{+0.047}_{-0.047}$& $0.872^{+0.065}_{-0.065}$& 0.799 &0.214& \citealp{Hasselfield13}\\
ACT, [BBN+H0+ACTcl(Dyn)] &SZ-mass & $0.301^{+0.082}_{-0.082}$& $0.975^{+0.108}_{-0.108}$&0.977 &0.207& \citealp{Hasselfield13}\\
SPT                     & SZ-mass & $0.276^{+0.047}_{-0.047}$& $0.781^{+0.037}_{-0.037}$&0.776&0.129& \citealp{Bocquet19}\\
HECS-SZ & SZ-mass&-- & -- &0.751&--&\citealp{Ntampaka19}\\
Planck18             & SZ-mass & $0.310^{+0.020}_{-0.020}$& $0.770^{+0.040}_{-0.040}$&0.783&0.138& \citealp{Zubeldia19}\\
\hline
\multicolumn{7}{c}{other cosmological probes}\\
\hline
DES-Y1                   & CS+GGL+AC$^{~(d)}$& $0.270^{+0.041}_{-0.040}$& $0.820^{+0.038}_{-0.036}$&0.778&0.143& \citealp{Abbott18b}\\
KiDS+GAMA           &CS+GGL+AC& $0.315^{+0.068}_{-0.092}$& $0.785^{+0.111}_{-0.117}$&0.804& 0.032&\citealp{Uitert18}\\
Pantheon            & SNe & $0.307^{+0.012}_{-0.012}$& ---&---& ---&\citealp{Scolnic18}\\
6dF+DR7+BOSS$^{~(e)}$   & BAO            & $0.346^{+0.045}_{-0.045}$& ---&---&---&\citealp{Alam17}\\

WMAP9          &  CMB only & $0.280^{+0.041}_{-0.040}$& $0.820^{+0.038}_{-0.036}$&0.792&0.112& \citealp{Hinshaw13}\\
Planck18            &  CMB only & $0.315^{+0.007}_{-0.007}$& $0.811^{+0.006}_{-0.006}$&0.832&0.000& \citealp{Planck18}\\
\hline
\end{tabular}
\begin{tablenotes}
\item {(a)} The cluster normalization condition parameter, $S_8$, is defined as $S_8 = \sigma_8 (\Omega_m/0.3)^{0.5}$ as used in the literature.
\item {(b)} $\Delta_{pl} = \sqrt{\left[(\Omega_{m,ref}-\Omega_{m,pl})/\Omega_{m,pl}\right]^2+\left[(\sigma_{8,ref}-\sigma_{8,pl})/\sigma_{8,pl}\right]^2}$ is the scatter of \om ~and \sig ~obtained from each method listed the table relative to that obtained from Planck18 \citep{Planck18}.
\item {(c)} \citet{Mantz15} used the combination of luminosity, temperature, gas mass, and lensing mass to estimate cluster mass which were refereed to as Weighting the Giant (WtG)
\item {(d)} CC = cosmic shear, GGL = galaxy-galaxy lensing, AC = angular clustering.
\item {(e)} 6dF = Six Degree Field Galaxy Survey \citep{Beutler11}, DR7 = SDSS data release 7 \citep{Ross15}, BOSS = Baryon Oscillation Spectroscopic Survey \citep{Alam17}
\end{tablenotes}
\end{table*}

\subsubsection{Selection Function Parameterization} \label{sec:selfunpara}
The constraints on \om ~and \sig ~is affected by parameterization of the selection function. Our selection function depends on three parameters $a, b$, and $\gamma$. The normalization $a$ is already fixed to unity. Assuming a normal distribution, the systematic of the selection function is investigated by generating $\sim 8000$ pairs of $b$ and $\gamma$, using the uncertainty in $b$ and $\gamma$ (see \S ~\ref{sec:SF}). For each pair we estimate the best fit values of \om ~and \sig. Figure \ref{fig:Fig06} shows the 68\% CL for the systematic of the selection function. This analysis rotates the error ellipses slightly compared to our fiducial analysis, but does not affect our results.
We obtain \om $= 0.313\pm0.035$ and \sig $= 0.809\pm0.012$, which is consistent with our result of the fiducial sample.

\subsection{Comparison with external data from cluster abundance} \label{sec:combine}
The left panel of Figure \ref{fig:Fig10} introduces the $68\%$ confidence level (CL) derived from $\mathtt{SelFMC}$ in comparison to the results obtained from other cluster abundance studies. Samples of galaxy cluster constructed from galaxy surveys include optical photometric (e.g., \citealp{Kirby19}), X-ray (e.g., \citealp{Mantz15}), and SZ (e.g., \citealp{Zubeldia19}) catalogs as listed in Table \ref{tab:Comp}. The figure shows that the CLs of all cluster abundance studies introduce a degeneracy between \om ~and \sig ~as we discussed in \S ~\ref{sec:const}. Also, the CL derived from $\mathtt{SelFMC}$ overlaps the CLs obtained from all other results as shown in the figure. Regardless of this overlapping, the right panel of Figure \ref{fig:Fig10} shows that the constraints on \om ~and \sig ~from cluster abundance studies are in tension with each other, even for the studies that use the same type of cluster sample. Specifically, the X-ray independent studies listed in Table \ref{tab:Comp} introduce different values of \om ~and \sig, which ~vary from $\sim$ 0.22 to 0.40 and 0.71 to 0.89, respectively. Also, the independent studies that use SZ-cluster samples show that \om ~and \sig ~vary from $\sim$ 0.25 to 0.31 and 0.77 to 0.98, respectively. 

The question is now, why are the cosmological constraints derived from many of the cluster abundance techniques in tension with each other? All cluster samples constructed from photometric surveys or detected by SZ effect do not return an estimate of each cluster's mass directly. For such samples the cluster mass has to be inferred indirectly  from other observables, which scale tightly with cluster mass. Among these mass proxies are X-ray luminosity, temperature, the product of X-ray temperature and gas mass (e.g., \citealp{Vikhlinin09,Mantz16a}), richness (e.g., \citealp{Yee03,Simet17}), and SZ signal (e.g., \citealp{Bocquet19}). To estimate cluster masses for the clusters in these samples it is necessary to follow up a subset of clusters and calculate their masses using, e.g., weak lensing or x-ray observations. Then, an observable-mass relation can be calibrated for these subsamples. Finally, the mass of each cluster in the sample can be estimated from this scaling relation. However, this reliance on observable-mass proxies introduces significant systematic uncertainties which is the dominant source of error (e.g., \citealp{Henry09,Mantz15}) for the reasons explained in the next paragraph. 

Firstly, the masses obtained for the follow-up subsample of clusters are often biased. For example, it is known that X-ray mass estimates are typically biased low and so a mass bias factor, (1-$\beta$), needs to be introduced and calibrated. Secondly, the size of the subsample used for calibration is usually small (tens of clusters) which introduces large uncertainties in both the slope and the normalization of the scaling relation. Thirdly, many cluster catalogs span a large redshift range so evolution (due to both the evolution of the universe and the physical processes of baryons in clusters) in the the scaling-relations used to estimate the masses needs to be carefully handled, introducing another source of uncertainty. All of the aforementioned assumptions can introduce large uncertainties in the estimates of cluster mass and consequently the constraints on cosmological parameters. For instance, \sig ~is specifically very sensitive to the high-mass end of the CMF and any offset of cluster true masses leads to biased estimation of \sig. Other observational systematics that introduce additional uncertainties are photometric redshift errors and cluster miscentering. 

\begin{figure*} \hspace*{0.25cm}
\includegraphics[width=20.5 cm]{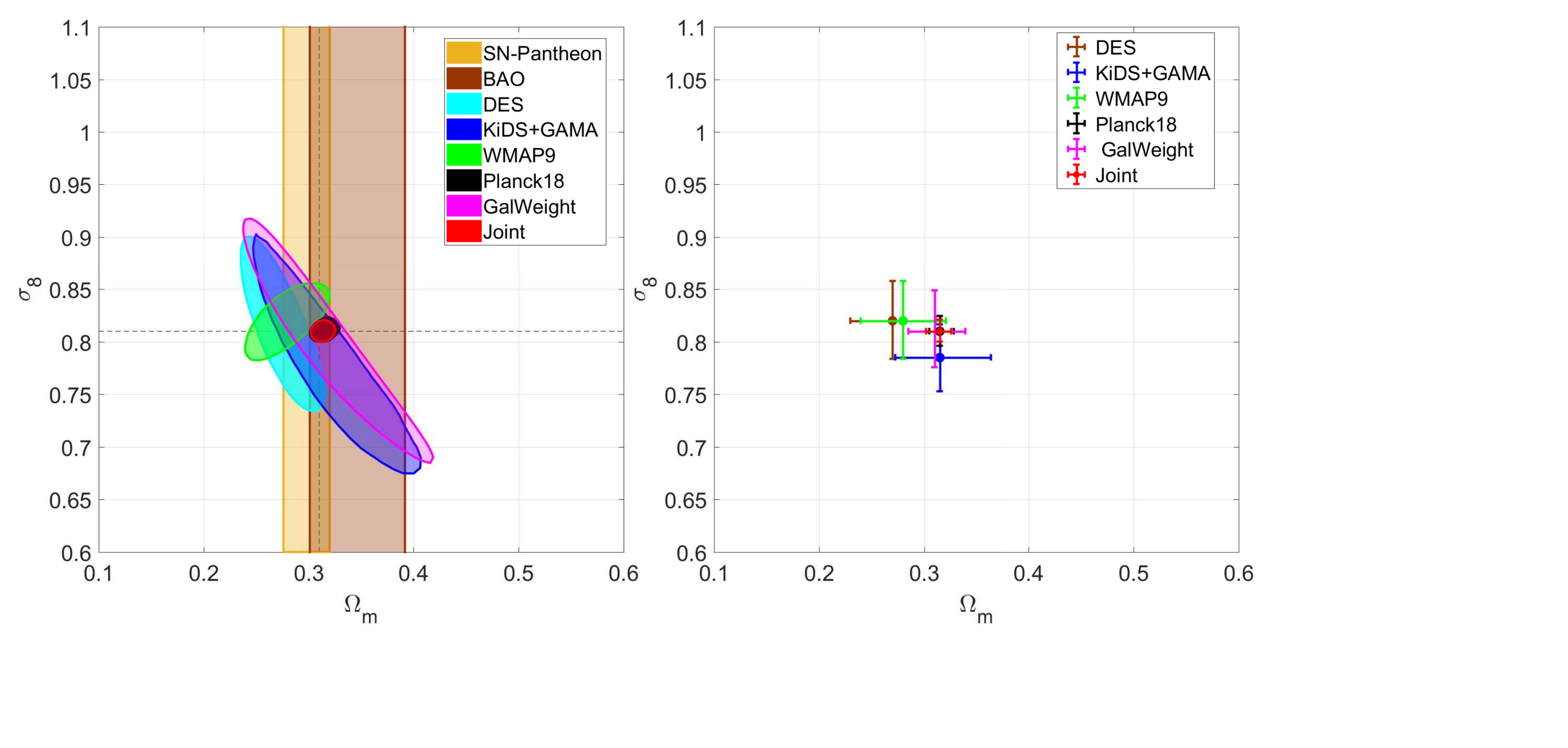} \vspace{-2.cm}
\caption{Constraints on \om ~and \sig ~obtained from cluster abundance ($\mathtt{SelFMC}$; magenta) and non-cluster cosmological constraint methods. Left: $68\%$ confidence levels (CLs) derived from $\mathtt{SelFMC}$, WMAP9 (CMB; \citealp{Hinshaw13}), Planck18 (CMB; \citealp{Planck18}), BAO data \citep{Beutler11,Ross15,Alam17}, Pantheon sample (SNe; \citealp{Scolnic18}), and the surveys KiDS+GAMA \citep{Uitert18} and DES Y1 \citep{Abbott18b} which both use the cosmological probes of cosmic shear, galaxy-galaxy lensing, and angular clustering. As in Figure \ref{fig:Fig10}, the two dashed lines show the best-fit values derived in this work. The constraints on \om ~and \sig ~derived from $\mathtt{SelFMC}$ are consistent with those derived from the non-cluster methods. Joint analysis between our constraints and the results of Planck18+BAO+Pantheon is represented by the red contour line. Right: uncertainties of \om ~and \sig ~estimated for the aforementioned probes except for the BAO and SNe probes which constrain \om ~only.} 
\label{fig:Fig11}
\end{figure*}

By using the $\mathtt{GalWCat19}$ cluster catalog and deriving cluster masses using the virial theorem, we were able to avoid most of the complexities described above. Firstly, we were able to identify clusters, assign membership, and determine cluster centers and  redshifts with high accuracy from the high-quality SDSS spectroscopic data set. Secondly, cluster membership was determined by the GalWeight technique which has been shown to be $\sim 98\%$ accurate in assigning cluster membership (Abdullah+18). Thirdly, a mass for each cluster was determined directly using the virial theorem. Therefore, we were able to recover a total (dark plus baryonic) mass for each cluster and circumvent having to make any assumptions about the complicated physical processes associated with the baryons. It has been suggested that cluster masses estimated via the virial theorem are overestimated by $~ 20\%$. But we note that we have applied a correction for the surface pressure term which we believe decreases this bias, especially when applied in combination with our GalWeight membership technique (Abdullah+18).
Abdullah+20 showed that the virial mass estimator performed well in comparison to the other mass estimators described in \citealp{Old15}, and resulted in a relatively low bias and scatter when applied to two semi-analytical simulations (see Figure 3 in Abdullah+20). Fourthly, since $\mathtt{GalWCat19}$ is a low-redshift cluster catalog it eliminates the need to make any assumptions about evolution in clusters themselves and evolution in cosmological parameters.  Finally, because of the large size of the $\mathtt{GalWCat19}$ we are able to determining the CMF well and consequently constrain the cosmological parameters \om ~and \sig\ with high precision. 

\subsection{Comparison with external data from non-cluster cosmological probes} \label{sec:combine1}
Cosmological parameters can be estimated from different cosmological probes rather than cluster abundance studies. We use measurements of primary CMB anisotropies from both WMAP (9-year data; \citealp{Hinshaw13}) and Planck satellites focused on the TT+lowTEB data combination from the 2018 analyses \citep{Planck18}. We also use angular diameter distances as probed by Baryon Acoustic Oscillations (BAO) including the 6dF Galaxy Survey \citep{Beutler11}, the SDSS Data Release 7 \citep{Ross15}, and the BOSS Data Release 12 \citep{Alam17}. Furthermore, we use measurements of luminosity distances from Type Ia supernovae from the Pantheon sample (\citealp{Scolnic18}). Finally, we use the measurements from a joint analysis of three cosmological probes: cosmic shear, galaxy-galaxy lensing, and angular clustering, including the results of the Kilo Degree Survey and the Galaxies And Mass Assembly survey (KiDS+GAMA; \citealp{Uitert18}) and the first year of the Dark Energy Survey (DES Y1; \citealp{Abbott18b}) (see Table \ref{tab:Comp}). The left panel of Figure \ref{fig:Fig11} introduces the $68\%$ CL derived from $\mathtt{SelFMC}$ in comparison to the those obtained from the aforementioned cosmological probes. As shown, the CL derived from $\mathtt{SelFMC}$ overlaps the CLs obtained from all non-cluster abundance probes. 

We define the scatter
\begin{equation}
\Delta_{pl} = \sqrt{\left(\frac{\Omega_{m,ref}-\Omega_{m,pl}}{\Omega_{m,pl}}\right)^2+\left(\frac{\sigma_{8,ref}-\sigma_{8,pl}}{\sigma_{8,pl}}\right)^2},
\end{equation}
to compare the constraints on \om ~and \sig ~obtained from all cosmological probes which are listed in Table \ref{tab:Comp} with that obtained from Planck18 \citep{Planck18}. Note that the constraints on \om ~and \sig ~derived from most of the cluster abundance studies independently introduce a relatively large scatter compared to the CMB experiment of Planck18. However, our constraints on \om ~and \sig ~are very comparable and competitive with Planck18 with a minimum value of $\Delta_{pl} = 0.018$. Moreover, our constraint on \om ~is in excellent agreement with the results of the BAO and Pantheon, separately. This remarkable consistency demonstrates that our derived cluster catalog at low redshift and calculating cluster masses using spectroscopic database of galaxy surveys is essential to obtain robust cosmological parameters. These results also emphasize the necessarily need to construct accurate cluster catalogs at high redshifts using the ongoing and upcoming galaxy surveys and perform similar analyses as introduced in this work. 

As discussed above there is a degeneracy between \om ~and \sig ~derived from the CMF at low redshift. We combine our $68\%$ CL with those obtained from Planck18+BAO+Pantheon, to eliminate the degeneracy of the our likelihood and to remarkably shrink the uncertainties of the cosmological parameters. The joint analysis gives \om ~$=0.315^{+0.013}_{-0.011}$  and \sig ~$=0.810^{+0.011}_{-0.01}$. 

\subsection{Conclusion} \label{sec:conc}
In this paper, we derived the CMF and the cosmological parameters \om ~and \sig ~using a mass-complete subsample of 756 clusters ($\mathtt{SelFMC}$) obtained from the $\mathtt{GalWCat19}$ cluster catalog which was constructed from SDSS-DR13 spectroscopic data set. The advantages of using this catalogs are: i) we were able to identify clusters, assign membership, and determine cluster centers and  redshifts with high accuracy from the high-quality SDSS spectroscopic data set; ii) cluster membership was determined by the GalWeight technique which has been shown to be $\sim 98\%$ accurate in assigning cluster membership (Abdullah+18); iii) the cluster masses were calculated individually using the virial theorem, and corrected for the surface pressure term; iv) $\mathtt{GalWCat19}$ is a low-redshift cluster catalog which eliminates the need to make any assumptions about evolution in clusters themselves and evolution in cosmological parameters; v) the size of $\mathtt{GalWCat19}$ is one of the largest available spectroscopic samples to be a fair representation of the cluster population.

Our CMF closely matches predictions from MultiDark Planck N-body simulations (snapshot hlist\_0.91520.list\footnote{\url{https://www.cosmosim.org/data/catalogs/NewMD_3840_Planck1/ROCKSTAR/trees/hlists/}}, with $z\sim 0.09$) for $\log(M)\gtrsim13.9$ \hm. Assuming a flat $\Lambda$CDM cosmology, we used the publicly available \verb|HMFcalc| \footnote{\url{http://hmf.icrar.org/}} code \citep{Murray13b} to estimate HMFs for the Tinker08 model (Equation \ref{eq:T08}). Then, using a standard $\chi^2$ procedure, we compared our cumulative mass function to HMFs to determine the best-fit mass function and constrain \om ~and \sig. We measured \om ~and \sig ~to be \om ~$=0.310^{+0.023}_{-0.027} \pm 0.041$ (systematic) and \sig ~$=0.810^{+0.031}_{-0.036}\pm 0.035$ (systematic), with a cluster normalization relation of $\sigma_8= 0.43 \Omega_m^{-0.55}$. 

The cosmological constraints we derived are very competitive with those recently derived using both cluster abundance studies and other cosmological probes. In particular, our constraint on \om ~and \sig ~are consistent with Planck18+BAO+Pantheon constraints. This remarkable consistency highlights the potential of using $\mathtt{GalWCat19}$ and its subsample $\mathtt{SelFMC}$ which are derived from SDSS-DR13 spectroscopic data set utilizing the application of GalWeight to produce precision constraints on cosmological parameters. The joint analysis of our cluster data with Planck18+BAO+Pantheon gives \om ~$=0.315^{+0.011}_{-0.013}$ and \sig ~$=0.810^{+0.011}_{-0.010}$.

\section*{ACKNOWLEDGMENTS}
We would like to thank Steven Murray for making his \verb|HMFcalc| calculator publicly available, and also for his guidance in running it. We also would like to thank Jeremy Tinker, Brian Siana, and Benjamin Forrest for their useful comments and help and Shadab Alam for providing us with the chain of BOSS-DR12 BAO data. Finally, we appreciate the comments and suggestions of the reviewer, which improved this paper. This work is supported by the National Science Foundation through grant AST-1517863, by HST program number GO-15294,  and by grant number 80NSSC17K0019 issued through the NASA Astrophysics Data Analysis Program (ADAP). Support for program number GO-15294 was provided by NASA through a grant from the Space Telescope Science Institute, which is operated by the Association of Universities for Research in Astronomy, Incorporated, under NASA contract NAS5-26555.

\begin{appendices}

\section{Evolution}\label{app:evo}

In this section, we discuss the evolution effect for a sample of clusters with a narrow redshift range between $z_1$ and $z_2$ with an average of $\langle z \rangle$. The HMF depends on the mass and redshift and is given by $\int^{z_2}_{z_1} n(M,z)dz/(z_2-z_1)$. 
We test the effect of evolution assuming an analytical model for the evolution of HMF and cosmological model with reasonable parameters. We then take the integral $\int^{z_2}_{z_1} n(M,z)dz/(z_2-z_1)$ and compare the results with $n(M,z)$ at $z = 0.085$. 

\begin{figure*} \hspace*{1.6cm}
\includegraphics[width=18cm]{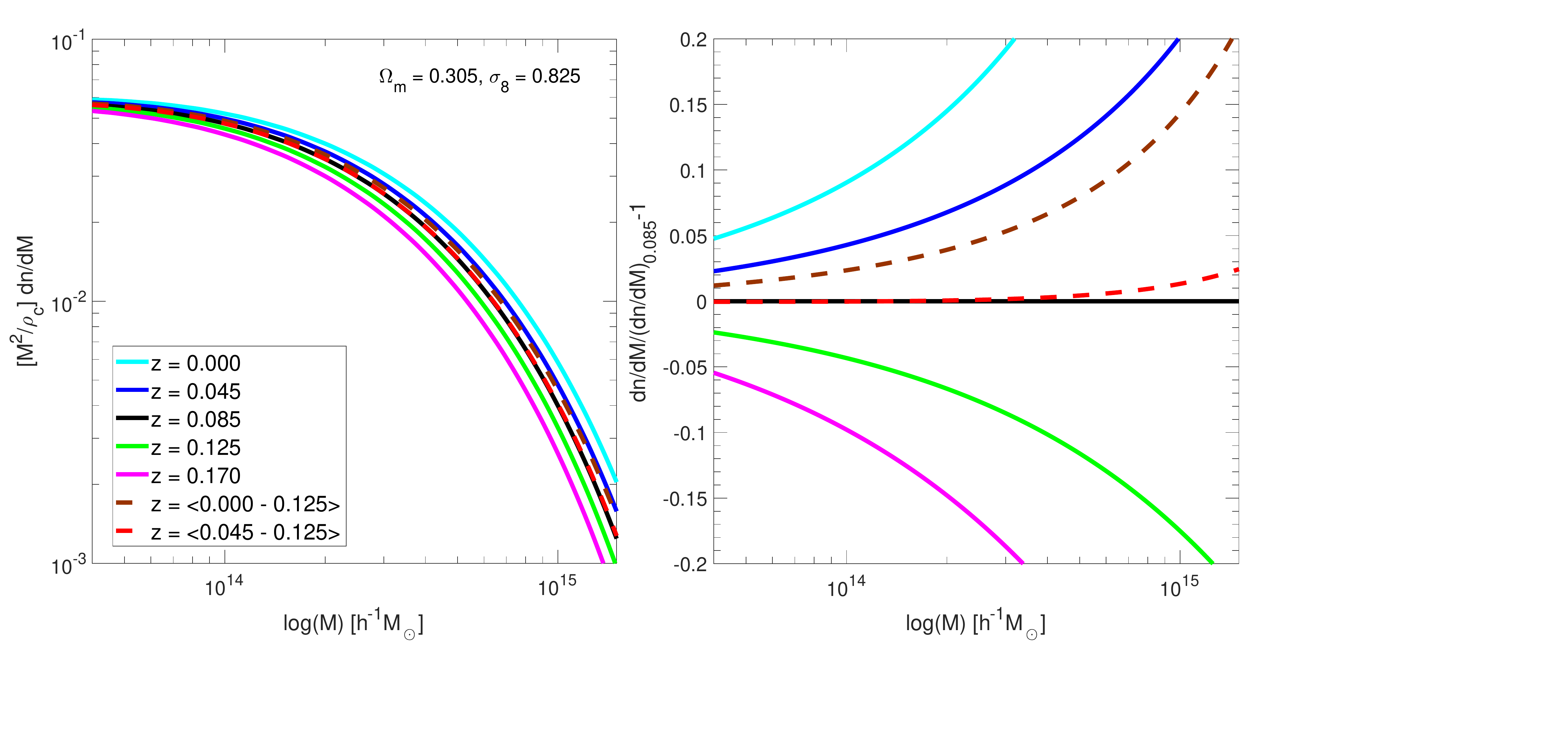} \vspace{-1.5cm}
\caption{The effect of cluster number density evolution. {\bf{Left}}: Tinker08 HMF times $M^2/\rho_c$ at different redshifts as well as the average HMF for $0.0\leq z \leq0.125$ (brawn) and $0.045\leq z \leq0.125$ (red) as shown in the legend. {\bf{Right}}: The scatter of each HMF relative to that at $z = 0.085$ (the mean redshift of the sample).
}
\label{fig:Fig08}
\end{figure*}

Figure \ref{fig:Fig08} shows the evolution of the cluster number density expected by Tinker08 for cosmological parameters \om = 0.305 and \sig = 0.825. In the left panel, we plot the HMF times $M/\rho_c$, $\rho_c$ is the critical density of the universe, to clarify the differences between the models at different redshifts. The right panel shows the scatter of models relative to the expectation at $z = 0.085$ (black line). As expected, the evolution of clusters with $z < 0.085$ is less than unity relative to that at $z = 0.085$ and the evolution of clusters with $z > 0.085$ is larger than unity relative to that at $z = 0.085$. The two dashed lines shows the expectation [$\int^{z_2}_{z_1} n(M,z)dz/(z_2-z_1)$] in the redshift intervals of  $0.0\leq z \leq0.125$ (brown) and $0.045\leq z \leq0.125$ (red). The plot indicates that the evolution is $>15\%$ for $0.0\leq z \leq0.125$ for massive clusters, while it drops to $<3\%$ for $0.045\leq z \leq0.125$.

Note that we do not neglect the effects of evolution. In other words, we do not assume that the HMF at $z_1$ is (nearly) the same as at $z_2$ (admittedly, there is 10-20\% difference in the most massive M). Because we use ratios of these quantities, most of the cosmological parameters (e.g., \sig) are canceled for sensible range (e.g., \sig = 0.75-0.85). We also test other HMF approximations such as Despali HMF \citep{Despali16} and obtain the same conclusion. Therefore, we restrict our data (observed clusters) to $0.045\leq z \leq0.125$ for which the  evolution effect of the number density of clusters is minimal.

\begin{figure*} \hspace*{1.25cm}
\includegraphics[width=36cm]{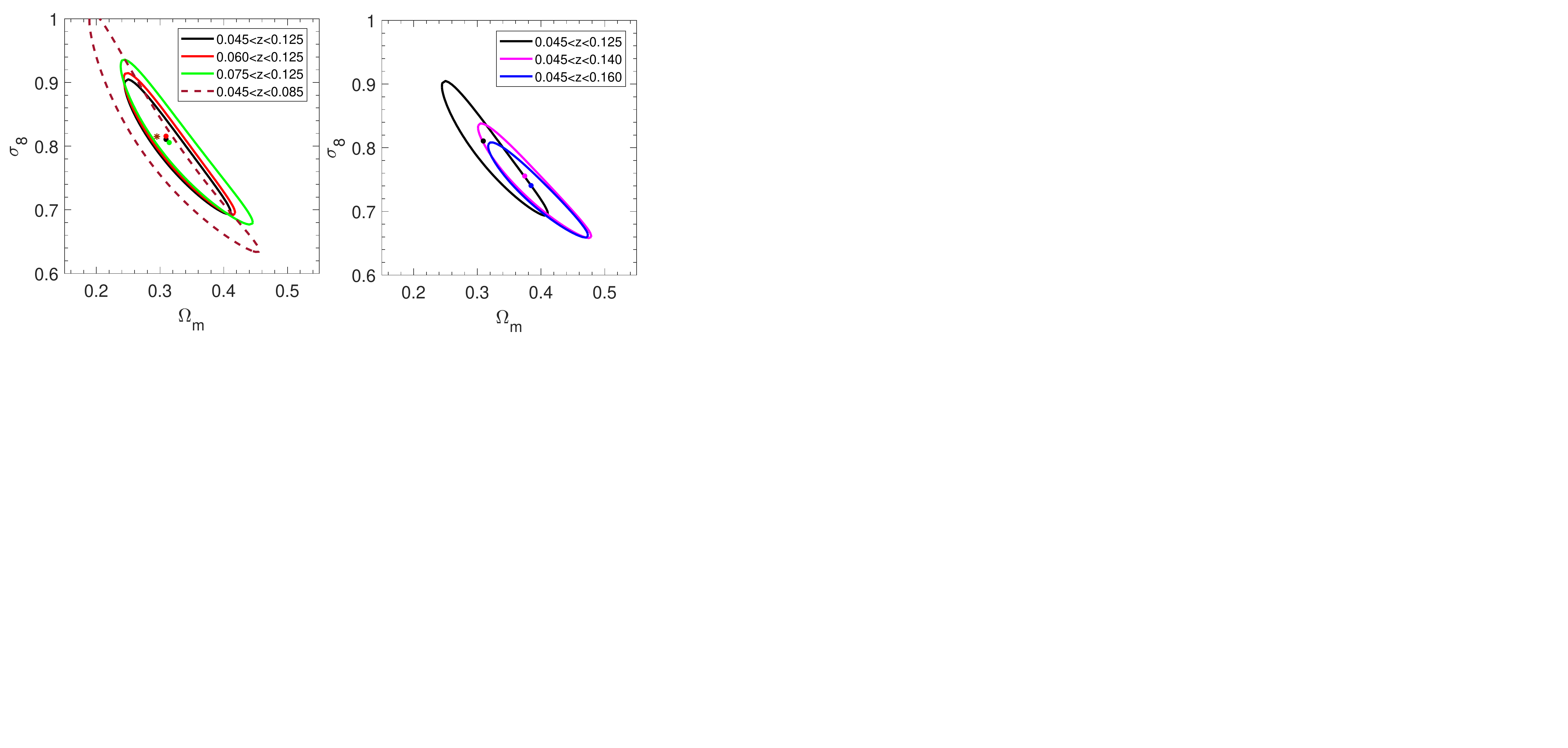} \vspace{-9.75cm}
\caption{The effect of adopting the redshift threshold. {\bf{Left}}: 68\% CLs for three subsamples with fixing the upper redshift threshold to 0.125 and decreasing the lower redshift threshold from 0.075 to 0.045. 
The dashed brown ellipse represents the 68\% CL of the $\mathtt{NoSelFVC}$ sample.
{\bf{Right}}: 68\% CLs for three subsamples with fixing the lower redshift threshold to 0.045 and increasing the upper redshift threshold from 0.125 to 0.16.}
\label{fig:Fig07}
\end{figure*}

\section{Redshift Threshold} \label{app:AA}

In this section we investigate the choice of the redshift interval and the application on the selection function of our results of the fiducial analysis as shown in Figure \ref{fig:Fig06}. In the left panel, we fix the upper redshift threshold to 0.125 and decrease the lower redshift threshold from 0.075 to 0.045. The plots indicates that decreasing the lower redshift threshold does not affect our result of the fiducial sample (black ellipse). It also demonstrates that the evolution effect is unremarkable in this small redshift interval. 
The left panel also introduces the 68\% CL of the $\mathtt{NoSelFVC}$ sample (dashed brown ellipse) which gives \om = $0.295^{+0.033}_{-0.034}$ (5\% less than the fiducial value) ~and \sig = $0.815^{+0.049}_{-0.050}$ (1\% greater than the fiducial value). The consistency between the results of $\mathtt{SelFMC}$ and $\mathtt{NoSelFVC}$ demonstrates that applying the selection function for $z\leq0.125$ does not affect the results of the fiducial analysis and is sufficient to correct for the volume incompleteness of $\mathtt{GalWCat19}$.
In the right panel, we fix the lower redshift threshold to 0.045 and increase the upper redshift threshold from 0.125 to 0.16. The plots indicates that increasing the upper redshift threshold significantly affects our constraints on \om ~and \sig ~because applying the selection function to higher redshift ($>0.125$) affects the shape of the CMF by increasing the scatter and noise and overcorrecting the number of clusters at high redshifts.

\end{appendices}

\bibliography{bibfile}

\end{document}